\documentclass[
 longbibliography,
 twocolumn,
 superscriptaddress,
 amsmath,
 amssymb,
 amsthm,
 amsfonts,
 pre,
 floatfix
]{revtex4-2}

\usepackage[colorlinks=true,urlcolor=blue,citecolor=blue,linkcolor=blue]{hyperref} 
\usepackage{graphicx}
\usepackage{dcolumn}
\usepackage{kotex}
\usepackage{natbib}
\usepackage{color}
\usepackage{amsfonts}
\usepackage{hyperref}
\usepackage{bm}

\newenvironment{psmallmatrix}
  {\left(\begin{smallmatrix}}
  {\end{smallmatrix}\right)}
  
\def\dbar{{\mathchar'26\mkern-12mu d}}
\DeclareMathOperator{\tr}{tr}

\newcommand{\RNum}[1]{\uppercase\expandafter{\romannumeral #1\relax}}

\begin{document}

\title{Inertial effects on the Brownian gyrator}

\author{Youngkyoung Bae}
 \affiliation{Department of Physics, Korea Advanced Institute of Science and Technology, Daejeon 34141, Korea}
 
\author{Sangyun Lee}
 \affiliation{Department of Physics, Korea Advanced Institute of Science and Technology, Daejeon 34141, Korea}

\author{Juin Kim}
\email{juinkim75@gmail.com}
\affiliation{Department of Physics and Chemistry,
Korea Air Force Academy, Cheongju, Chungbuk 28187, Korea}

\author{Hawoong Jeong}
\email{hjeong@kaist.edu}
\affiliation{Department of Physics, Korea Advanced Institute of Science and Technology, Daejeon 34141, Korea}
\affiliation{Center for Complex systems, Korea Advanced Institute of Science and Technology, Daejeon 34141, Korea}

\date{\today}

\begin{abstract}
The recent interest into the Brownian gyrator has been confined chiefly to the analysis of Brownian dynamics  both in theory and experiment despite the applicability of general cases with definite mass. Considering mass explicitly in the solution of the Fokker--Planck equation and Langevin dynamics simulations, we investigate how inertia can change the dynamics and energetics of the Brownian gyrator. In the Langevin model, the inertia reduces the nonequilibrium effects by diminishing the declination of the probability density function and the mean of a specific angular momentum, $j_\theta$, as a measure of rotation. Another unique feature of the Langevin description is that rotation is maximized at a particular anisotropy while the stability of the rotation is minimized at a particular anisotropy or mass. Our results suggest that the Langevin dynamics description of the Brownian gyrator is intrinsically different from that with Brownian dynamics. In addition, $j_\theta$ is proven to be essential and convenient for estimating stochastic energetics such as heat currents and entropy production even in the underdamped regime. 
\end{abstract}

\maketitle

\section{Introduction}
\label{sec:intro}

On account of its simplicity and efficiency, Brownian dynamics has been adopted in a series of recent studies to describe biological systems such as chromosomes~\cite{weber2012nonthermal}, primary cilia~\cite{battle2015intracellular, battle2016broken}, membrane fluctuations~\cite{gov2004membrane, ben2011effective}, and actin-myosin networks~\cite{mizuno2007nonequilibrium, gladrow2016broken, gladrow2017nonequilibrium}. In many cases, characteristic directed currents in configuration space reveal the violation of detailed balance originating from thermal nonequilibrium (see the rotational probability currents in steady state in Refs.~\cite{lander2012noninvasive, battle2016broken, gnesotto2018broken}). Studies of such biological nonequilibrium systems through Brownian dynamics have expanded our understanding of fluctuation-dissipation theorem~\cite{baiesi2009fluctuations, sharma2016communication, asheichyk2019response}, fluctuation theorems~\cite{gallavotti1995dynamical, kurchan1998fluctuation, seifert2005entropy, seifert2012stochastic} and the thermodynamic uncertainty relation~\cite{barato2015thermodynamic, gingrich2016dissipation, pietzonka2017finite, pietzonka2018universal, chun2019effect}. 

The choice of Brownian dynamics may be appropriate in describing such systems because it reduces simulation cost when long-time configurational dynamics are the main interest and short-time movements do not change the results significantly. However, the development of the related theory and experiments is moving our concern to faster motions that could result in crucial differences. Observation of short-time dynamics have become available at greater time resolutions so that we are able to examine a number of theoretical results based on Langevin dynamics, where the memory effect caused by the inertia of a particle is relevant~\cite{blum2006measurement, li2010measurement, huang2011direct, pusey2011brownian}. Moreover, systems in low-density environments (e.g., rarefied gas~\cite{blum2006measurement}) or at large scales such as flocks of birds~\cite{attanasi2014information}, schools of fish~\cite{katz2011inferring}, vibrobots~\cite{giomi2013swarming}, and various mesoscale organisms~\cite{selmeczi2005cell, rabault2019curving, klotsa2019above} should be addressed by Langevin dynamics including the inertial term to more realistically catch their characteristics. Normally, Langevin dynamics correspond to larger masses, lower frictions, and shorter time scales compared to Brownian dynamics.

Though Brownian dynamics is an overdamped limit of Langevin dynamics, neglecting the inertial term is not always successful even in longer time scales. There have been reports that the overdamped approximation fails in a spatially inhomogeneous temperature field~\cite{benjamin2008inertial, celani2012anomalous} or in the presence of a magnetic field~\cite{ao2007existence, yuan2017sde, Chun2018Emergence, Lee2019Nonequilibrium}.
While studies to explain the inertial effects have shown that inertia qualitatively changes the system dynamics of a motility-induced phase separation~\cite{mandal2019motility} as well as the dynamical states and translational motion of a self-propelled particle~\cite{scholz2018inertial, dauchot2019dynamics, gutierrez2020inertial, lowen2020inertial, caprini2020inertial}, the effects of inertia on rotational motion and system energetics have been less considered; thus, how inertia affects the dynamics and energetics of a wider range of nonequilibrium systems, including the Brownian gyrator, still remains unclear. 

In this paper, we investigate inertial effects on the dynamics and energetics of the Brownian gyrator~\cite{filliger2007brownian}, which is a two-dimensional (2D) model treating the rotational motion of a particle in contact with two different heat baths and in an anisotropic harmonic potential. This model is widely used not only because it is exactly solvable but also because it can be interpreted as a bead-spring model of an internally driven assembly in biological systems~\cite{battle2016broken, mura2018nonequilibrium, Gradziuk2019Scaling}. However, the absence of the inertial term in Brownian dynamics bears some critical limitations.
First, even though the concept of the Brownian gyrator has been realized (as the overdamped limit) in recent experiments with stochastic electronic and colloidal systems~\cite{cilberto2013heat, ghanta2017fluctuation, argun2017experimental, chiang2017electrical, gonzalez2019experimental}, it is still possible to further develop the idea to more general experiments where the particle has considerable mass. In that case, there is lack of research with which to compare the results.
Regarding the rotational motion of a particle in nonequilibrium steady state (NESS), a curl of probability currents and a cycling frequency of the Brownian gyrator has been studied~\cite{dotsenko2013two, cerasoli2018asymmetry, mancois2018two, nascimento2019memory}, but most related reports have not considered particle inertia. In this respect, including the inertial term, i.e., adopting Langevin dynamics, will be beneficial for clarifying the actual rotational motion of a particle in NESS. Our results here reveal that consideration of inertia remarkably changes the probability density of the particle and its rotational motion. Further, we derive the relation between the energetics and the rotational motion in the underdamped regime and show that energetic quantities can be inferred from dynamical properties. 

This paper is organized as follows. Section~\ref{sec:2} introduces the Brownian gyrator and its nonequilibrium features through Brownian dynamics, which we call the overdamped model. Section~\ref{sec:3} describes how the inertial term in Langevin dynamics changes the system dynamics mainly concerning the rotational motion, which we call the inertial model. Section~\ref{sec:4} clarifies how the stochastic energetics relates to rotational motion in the underdamped regime. 

\section{Overdamped Model}
\label{sec:2}

\subsection{Tilted PDF and rotational motion} 
\begin{figure}[t]
    \includegraphics[width=\columnwidth]{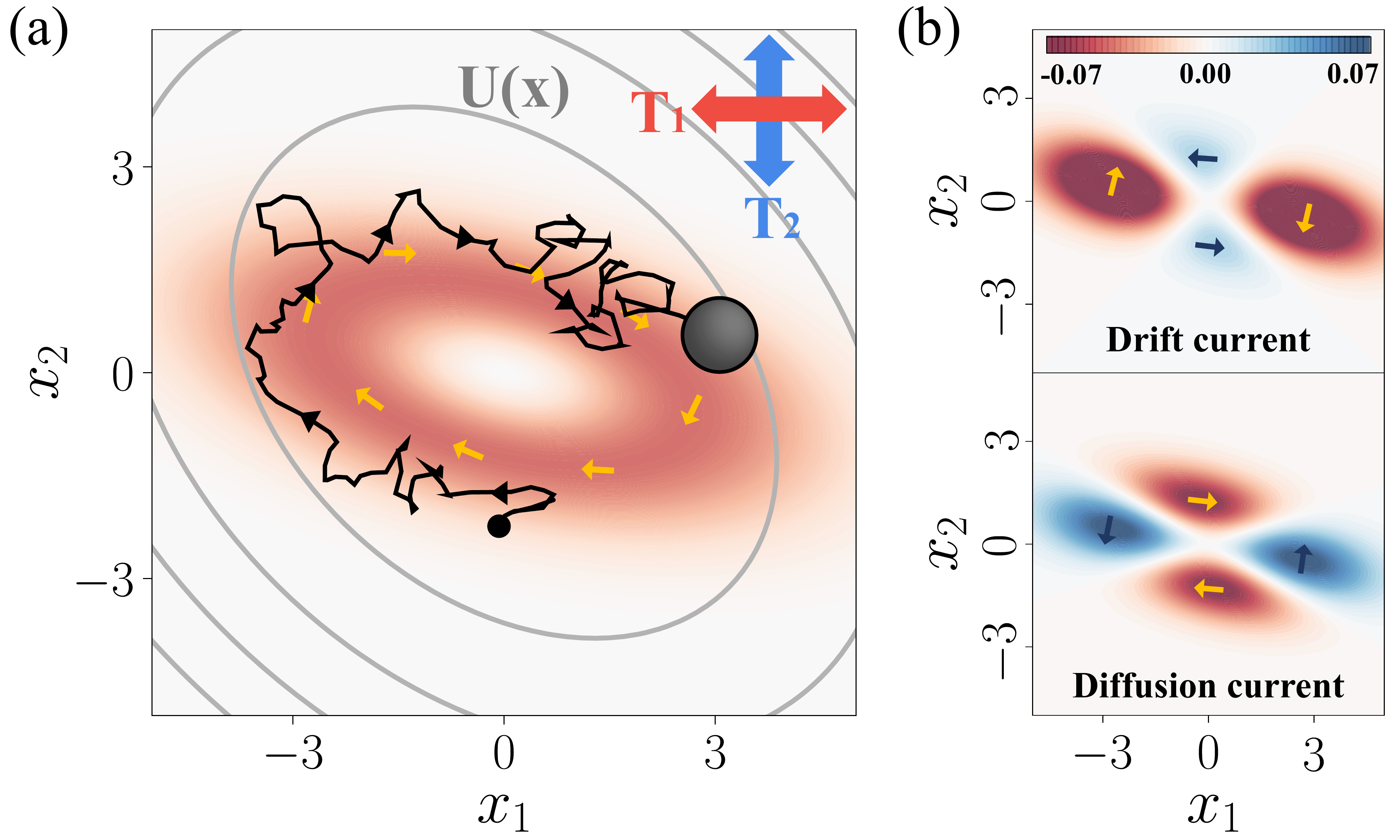}
    \vskip -0.1in
    \caption{ (a) Schematic diagram of a Brownian gyrator with an angular current $\bm{x} \times \bm{j}_{\bm{x}}(\bm{x})$ (colored contour). An anisotropic harmonic potential $U(\bm{x})$ is shown as gray contour lines, and the black line indicates a numerically generated trajectory of the particle. (b) Top and bottom panels show angular current contributions of the drift and diffusion currents, respectively. By adding these two currents, the particle undergoes a rotational motion in a 2D plane. A positive angular current represents clockwise rotation, and the small arrows indicate the local directions of the averaged currents. A color-bar of colored contours is given in the top panel.
    The parameters are fixed as $k=3/2$, $u=1/2$, $T_1 = 5$, $T_2 = 1$, and $\gamma = 1$. }\label{fig1}
\end{figure}

In the overdamped model, we consider a particle moving in a 2D plane with the position $\bm{x} \equiv (x_1, x_2)^T$ and neglect the inertial term. The particle undergoes an anisotropic harmonic potential, $U(\bm{x}) = \frac{1}{2}\bm{x}^T \cdot \mathsf{U} \cdot \bm{x}$ where $\mathsf{U} = \begin{psmallmatrix} k & u  \\ u & k \end{psmallmatrix}$ with $u < k$, and it contacts with two different heat baths at temperature $T_1$ and $T_2$ ($< T_1$); see Fig. \ref{fig1}(a). Then, the Langevin equation for this model can be written as 
\begin{equation}\label{eq:OverLangevinEq}
\begin{aligned}
    \gamma \dot{\bm{x}}(t) &= -\nabla_{\bm{x}} U(\bm{x}(t)) + \bm{\xi}(t),
\end{aligned}
\end{equation}
where $\gamma$ is the Stokes friction coefficient and $\bm{\xi} \equiv (\xi_1, \xi_2)^T$ is a Gaussian white noise satisfying $\langle \xi_i(t)\rangle=0$ and $\langle \xi_i(t)\xi_j(t') \rangle = 2\gamma T_i\delta_{ij} \delta(t-t')$. The angle bracket $\langle \cdot \rangle$ stands for the ensemble average. We set Boltzmann's constant $k_B = 1$ and all parameters are dimensionless. 
The anisotropic potential and the different heat baths may be equivalently thought of as a simple shear flow~\cite{asheichyk2019using} and an additional Gaussian white noise in one direction~\cite{argun2017experimental}.

%%%%%%%%%%%%% figure 1 %%%%%%%%%%%%%%
%%%%%%%%%%%%% figure 1 %%%%%%%%%%%%%%

%%%%%%%%%%%%% figure 2 %%%%%%%%%%%%%%
\begin{figure*}[ht]
    \includegraphics[width=\linewidth]{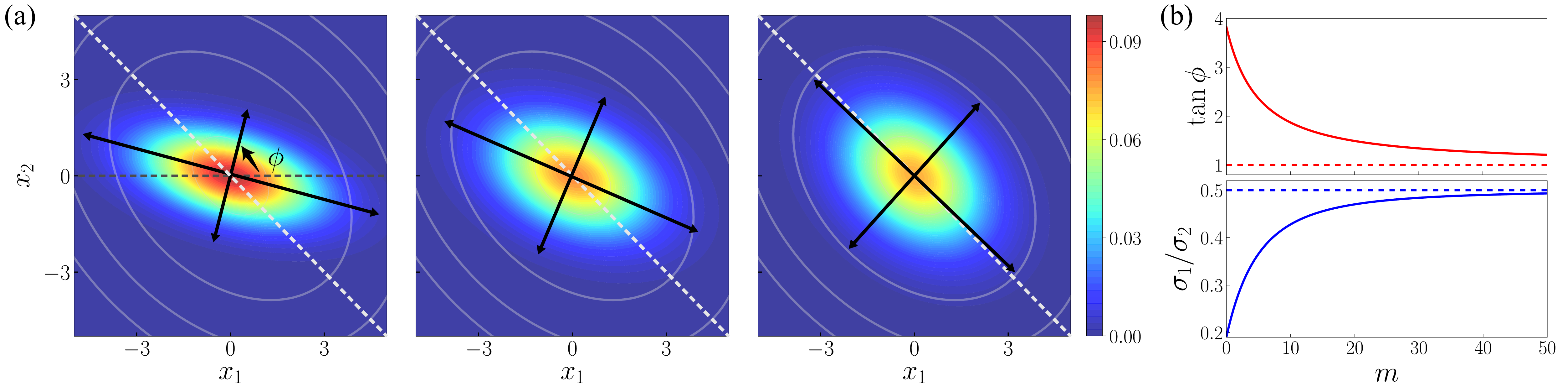}
    \vskip -0.1in
    \caption{(a) Positional PDFs $p(\bm{x})$ with mass $m=0$, $5$, and $50$ from left to right. A harmonic potential $U(\bm{x})$ is shown as gray contour lines and the PDFs are plotted as colored contours. Here, the principal axes of the PDFs and potential $U(\bm{x})$ are indicated by black arrows and white dotted lines, respectively. (b) Analytical results of the slope of the minor principal axis ($\tan \phi$) and the aspect ratio of the variances along the principal axes ($\sigma_1/\sigma_2$) of the PDF as a function of mass $m$. Dotted lines indicate the asymptotic lines in the limit of $m \to \infty$. The other parameters are fixed as $k=3/2$, $u=1/2$, $T_1 = 5$, $T_2 = 1$, and $\gamma = 1$.}\label{fig2}
\end{figure*}
%%%%%%%%%%%%% figure 2 %%%%%%%%%%%%%%

To obtain the probability density function (PDF) and the probability current, we consider the associated Fokker--Planck equation given by
\begin{equation}\label{eq:OverFPEq}
\begin{aligned}
    \frac{\partial p(\bm{x}, t)}{\partial t} = -\nabla_{\bm{x}} \cdot \bm{j}_{\bm{x}}(\bm{x}, t),
\end{aligned}
\end{equation}
where the probability current $\bm{j}_{\bm{x}}(\bm{x}, t)$ is defined by
\begin{equation}\label{eq:OverCurrent}
\begin{aligned}
    \bm{j}_{\bm{x}} (\bm{x}, t) = -\left[ \frac{1}{\gamma} \nabla_{\bm{x}} U(\bm{x})+ \mathsf{D}\cdot \nabla_{\bm{x}} \right] p(\bm{x}, t).
\end{aligned}
\end{equation}
Here, the diffusion matrix is given as $\mathsf{D} \equiv \frac{1}{\gamma}\begin{psmallmatrix} T_1 & 0 \\ 0 & T_2 \end{psmallmatrix}$. The first term on the right-hand side of Eq.~\eqref{eq:OverCurrent} is the drift current determined by potential $U(\bm x)$, and the second term is the diffusion current of the system~\cite{risken1996fokker}.

We calculate the steady-state PDF $p(\bm{x})$ using the method in Appendix \ref{sec:AppendixA}.
The covariance matrix in the steady-state $\mathsf{C}_{\bm{x}\bm{x}}$ for $\bm{x}$ defined as $\langle \bm{x}\bm{x}^T \rangle$ is given by
\begin{equation}\label{eq:OverCorr}
\begin{aligned}
    \mathsf{C}_{\bm{x}\bm{x}} = \frac{1}{2k(k^2 - u^2)}
        \begin{psmallmatrix}
        2 T_1 k^2 + (T_2 - T_1)u^2 & -(T_1 + T_2)ku  \\
        -(T_1 + T_2)ku & 2 T_2 k^2 + (T_1 - T_2)u^2
        \end{psmallmatrix},
\end{aligned}
\end{equation}
and $p(\bm{x})$ is
\begin{equation}\label{eq:OverDensity}
\begin{aligned}
    p(\bm{x}) = \frac{1}{2 \pi \sqrt{\det{\mathsf{C}_{\bm{x}\bm{x}}}}} \exp{ \left( -\frac{1}{2} \bm{x}^T \cdot \mathsf{C}_{\bm{x}\bm{x}}^{-1} \cdot \bm{x} \right)}.
\end{aligned}
\end{equation}
Inserting $p(\bm{x})$ into Eq.~\eqref{eq:OverCurrent}, we can easily obtain the probability current $\bm{j}_{\bm{x}}(\bm{x})$. When $T_1 \neq T_2$ and $u \neq 0$, $\bm{j}_{\bm{x}}(\bm{x})$ has a non-zero value, which implies that the system is in NESS. The rotational property of $\bm{j}_{\bm{x}}(\bm{x})$ is represented as an angular current, denoted as $\bm{x}\times \bm{j}_{\bm{x}}(\bm{x})$ in Fig.~\ref{fig1}(a), where $\bm{x} \times \bm{j}_{\bm{x}}(\bm{x}) \equiv x_1 j_{x, 2}(\bm{x}) - x_2 j_{x, 1}(\bm{x})$. This current can be divided into drift and diffusion parts as seen in Fig.~\ref{fig1}(b); hence, we demonstrate that the resulting rotational motion arises from the combined effects of these two angular currents.

One of the features of NESS is a tilted PDF compared to the equilibrium state whose shape is determined by a potential.
For example, when our system is in equilibrium ($T_1 = T_2 = T$), the inverse of the covariance matrix is given by $\mathsf{C}_{\bm{x}\bm{x}}^{-1} = \frac{1}{T} \mathsf{U}$.
In this case, the principal axes of the PDF and the aspect ratio of the variances along the principal axes ($\sigma_1/\sigma_2$) coincide with the values of potential $\mathsf{U}$. However, in NESS, the PDF is tilted to a higher temperature axis (i.e., $x_1$--axis) compared to the equilibrium state, as shown in Fig.~\ref{fig2}(a), and thus the PDF cannot fully cover the potential. In other words, the principal axes of $\mathsf{C}_{xx}^{-1}$ do not coincide with the principal axes of $\mathsf U$ in NESS.

The NESS is characterized by a non-zero probability current that rotates around the center. To quantify this rotational motion, we set the specific angular momentum as $j_\theta (\bm{x}, t) \equiv \bm{x}(t) \times \bm{\nu}(\bm{x}, t) $, where the mean local velocity conditioned on $\bm{x}$ is $\bm{\nu}(\bm{x}, t) \equiv \langle \dot{\bm{x}} | \bm{x}, t \rangle = \bm{j}_{\bm{x}}(\bm{x}, t) / p(\bm{x}, t)$~\cite{seifert2012stochastic}. This term,
$j_\theta(\bm{x}, t)$, is related to the stochastic area tensor~\cite{ghanta2017fluctuation, gonzalez2019experimental} as well as the probability angular momentum~\cite{weiss2019nonequilibrium}, which have been proposed as measures of the violation of detailed balance.
The mean of specific angular momentum $\langle j_\theta \rangle_{ss}$ can be evaluated as
\begin{equation}\label{eq:OverSAM}
\begin{aligned}
    \langle j_\theta \rangle_{ss} 
    &= \langle \bm{x} \times \bm{\nu}(\bm{x}, t) \rangle_{ss} = \frac{u}{k \gamma} (T_2 - T_1),
\end{aligned}
\end{equation}
where $\langle \cdot \rangle_{ss}$ denotes the ensemble average in the steady state.
In Eq.~\eqref{eq:OverSAM}, $\langle j_\theta  \rangle_{ss}$ is proportional to $u (T_2 - T_1)$, which means that the rotational motion is caused by two effects: temperature difference and anisotropy of the potential. 
Since the rotational motion of a particle reflects that the system is in NESS, we can confirm that two different temperatures and an anisotropy of potential are the sources of the nonequilibrium state.

\section{Inertial Model}
\label{sec:3}
\subsection{Steady-state PDF: Approaching equilibrium}
To investigate the effects of inertia on a particle in NESS, we consider the Brownian gyrator in an underdamped regime, called the inertial model.
The Langevin equation of a particle of mass $m$ is given by
\begin{equation}\label{eq:InertialLangevinEq}
\begin{aligned}
    \dot{\bm{x}}(t) &= \bm{v}(t),\\
    m\dot{\bm{v}}(t) &= -\nabla_{\bm{x}} U(\bm{x}(t)) -\gamma \bm{v}(t)  + \bm{\xi}(t),
\end{aligned}
\end{equation}
where the velocity $\bm{v} \equiv (v_1, v_2)^T$. The PDF $p(\bm{x}, \bm{v}, t)$ in the inertial model satisfies the Fokker--Planck equation associated with Eq.~\eqref{eq:InertialLangevinEq} written as
\begin{equation}\label{eq:FPeq}
\begin{aligned}
    \frac{\partial p(\bm{x}, \bm{v}, t)}{\partial t} &= -\nabla_{\bm{x}} \cdot \bm{j}_{\bm{x}} (\bm{x}, \bm{v}, t) - \nabla_{\bm{v}} \cdot \bm{j}_{\bm{v}} (\bm{x}, \bm{v}, t),
\end{aligned}
\end{equation}
where the probability currents are given as
\begin{equation}\label{eq:InertialCurrents}
\begin{aligned}
    \bm{j}_{\bm{x}} (\bm{x}, \bm{v}, t) &= \bm{v} p(\bm{x}, \bm{v}, t), \\
    \bm{j}_{\bm{v}} (\bm{x}, \bm{v}, t) &= -\left(\frac{1}{m}\mathsf{U} \cdot \bm{x} + \frac{1}{m}\mathsf{\Gamma} \cdot \bm{v} + \mathsf{D} \cdot \nabla_{\bm{v}}\right) p(\bm{x}, \bm{v}, t).
\end{aligned}
\end{equation}
Here, $\mathsf{\Gamma} \equiv \begin{psmallmatrix} \gamma & 0 \\ 0 & \gamma \end{psmallmatrix}$ and $\mathsf{D} \equiv \frac{\gamma}{m^2}\begin{psmallmatrix} T_1 & 0 \\ 0 & T_2 \end{psmallmatrix}$ are $2 \times 2$ matrices related to the dissipation due to friction and the diffusion, respectively. The steady-state PDF $p(\bm{z})$ is obtained as 
\begin{equation}\label{eq:InertialDensity}
\begin{aligned}
    p(\bm{z}) = \frac{1}{2 \pi \sqrt{\det{\mathsf{C}}}} \exp{ \left( -\frac{1}{2} \bm{z}^T \cdot \mathsf{C}^{-1} \cdot \bm{z} \right)},
\end{aligned}
\end{equation}
where the state vector $\bm{z} \equiv (x_1, x_2, v_1, v_2)^T$ and the covariance matrix $\mathsf{C} \equiv \langle \bm{z}\bm{z}^T \rangle_{ss}$. The complete expression of $\mathsf{C}$ is given in Appendix~\ref{sec:AppendixA}.

Integrating the PDF of Eq.~\eqref{eq:InertialDensity} over $\bm{v}$, the change of positional PDF with $m$ is illustrated in Fig.~\ref{fig2}(a). Compared with the overdamped model (leftmost), the PDF with inertia is less tilted, and its elliptical shape becomes more circular as $m$ increases. To quantify this asymptotic behavior of the PDF, we take two measures: the slope of the minor principal axis of the PDF ($\tan \phi$) and the aspect ratio of the variances along the principal axes ($\sigma_1/\sigma_2$). In evaluating $\tan \phi$, $\phi$ is the tilt angle between the minor principal axis and the $x_1$-axis, as shown in Fig.~\ref{fig2}(b). The variances along the principal axes denoted as $\sigma_1$ and $\sigma_2$ ($> \sigma_1$) are obtained by the eigenvalues of the covariance matrix of $\bm{x}$. 

Figure~\ref{fig2}(b) plots the analytical results of the $\tan \phi$ and $\sigma_1/\sigma_2$ measures as a function of $m$. They converge to $1$ and $(k-u)/(k+u)$ in the limit of $m \to \infty$, respectively, where the convergent values are equal to the values of potential $\mathsf{U}$. From the fact that the tilted PDF is one piece of evidence for the NESS, these reflect that the system approaches equilibrium from the NESS as the inertia becomes significant. This result is not surprising because the diffusion matrix $\mathsf{D}$ is inversely proportional to $m^2$, and hence the diffusion effects caused by the temperature difference are also diminished with $m$.

%%%%%%%%%%%%% figure 3 %%%%%%%%%%%%%%
\begin{figure}[t]
    \centering
    \includegraphics[width=\columnwidth]{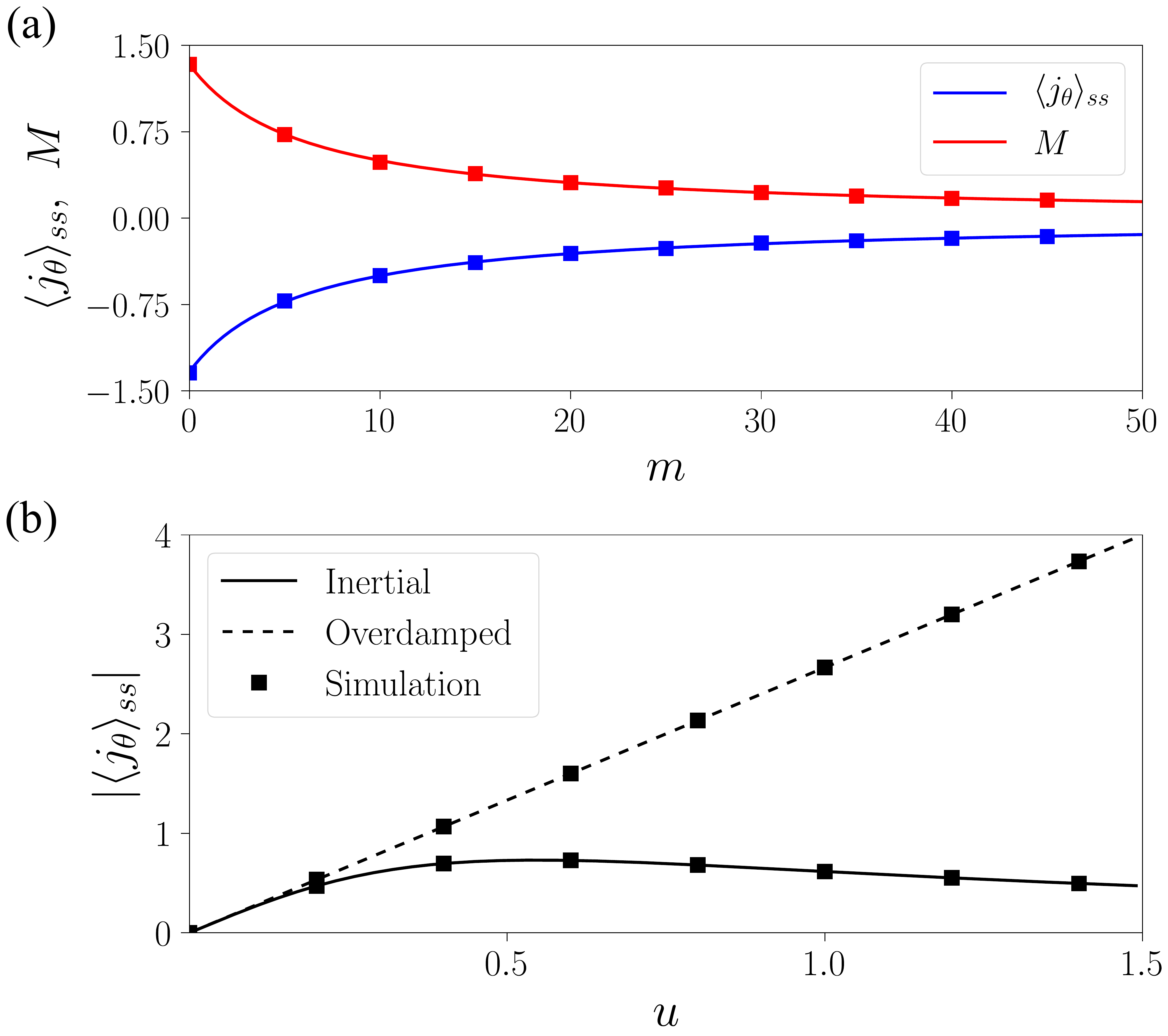}
    \vskip -0.1in
    \caption{(a) Systematic torque $M$ (upper) and the mean of specific angular momentum $\langle j_\theta \rangle_{ss}$ (lower) as a function of mass $m$ with $u=1/2$. Both vanish in the limit $m \to \infty$. (b) $|\langle j_\theta \rangle_{ss} |$ as a function of $u$. In both figures, solid (dotted) lines represent the analytical results of the inertial model with $m=5$ (the overdamped model). Squares represent Langevin simulation results. The parameters are fixed as $k=3/2$, $T_1 = 5$, $T_2 = 1$, and $\gamma = 1$.}\label{fig3}
\end{figure}
%%%%%%%%%%%%% figure 3 %%%%%%%%%%%%%%

\subsection
{Specific angular momentum and its fluctuation: Non-monotonic behaviors}

In the inertial model, the specific angular momentum is defined as $j_\theta(\bm{x}, \bm{v}, t) \equiv \bm{x}(t) \times \bm{v}(t)$. The mean, $\langle j_\theta \rangle_{ss}$, can be obtained using the covariance between $\bm{x}$ and $\bm{v}$ in $\mathsf{C}$, but it is also possible using the relation with a systematic torque $M$ which the particle exerts on the potential as
\begin{equation}\label{eq:Torque&SAM}
    \langle j_\theta \rangle_{ss} = -\frac{1}{\gamma} M,
\end{equation}
where $M \equiv \langle \bm{x} \times \nabla_{\bm{x}} U(\bm{x}) \rangle_{ss}$~\cite{filliger2007brownian}. This relation has been shown in the overdamped system in Ref.~\cite{mura2018nonequilibrium}, but a proof with inertia has been absent.
We derive this relation in the inertial model in Appendix~\ref{sec:Appendix_Torque&SAM}. Using Eq.~\eqref{eq:Torque&SAM}, $\langle j_\theta \rangle_{ss}$ in the inertial model is obtained as
\begin{equation}\label{eq:InertialSAM}
\begin{aligned}
    \langle j_\theta \rangle_{ss} = -\frac{1}{\gamma}M = \frac{u}{\gamma} \langle x_2^2 - x_1^2 \rangle_{ss} = \frac{\gamma u (T_2 - T_1)}{k\gamma^2 + u^2 m}.
\end{aligned}
\end{equation}
Here, $\langle j_\theta \rangle_{ss}$ is determined by the difference between the two variances, i.e. how much the positional PDF is tilted. Thus, the tilted PDF and the rotational motion of the particle are not separate but highly related phenomena. This relation gives us not only a simpler calculation but an important feature for measuring $\langle j_\theta \rangle_{ss}$.
In the underdamped regime, measuring the velocity field is necessary for a full description of the Langevin dynamics. Nevertheless, $\langle j_\theta \rangle_{ss}$ can be estimated by experimentally given quantities and measurements of the variances of displacements from a positional trajectory of a particle without knowledge of the velocity field.

We take into account two parameters, $m$ and $u$, to describe how and to what extent the system dynamics are changed with the inertial term since $u$ is a controllable parameter in experiments and has a crucial role in keeping the NESS in our system.
Figure~\ref{fig3}(a) represents $\langle j_\theta \rangle_{ss}$ and $M$ as functions of $m$ and shows that they vanish as $m$ increases, similarly with the tilted PDF.
The vanishing behavior of $\langle j_\theta \rangle_{ss}$ has been revealed in the form of a decreasing cycling frequency with increasing $m$ in Ref.~\cite{mancois2018two}.
This result can be intuitively understood by considering the ensemble-averaged moment of inertia of the particle, $I_\theta$, which is derived as
\begin{equation}\label{eq:MomInertia}
\begin{aligned}
    I_\theta \equiv m \langle x_1^2 + x_2^2 \rangle_{ss} = m \frac{k(T_1 + T_2)}{k^2 - u^2}.
\end{aligned}
\end{equation}
$I_\theta$ is directly proportional to $m$ as expected; accordingly, finding the rotational motion is more difficult for large $m$.
The interesting point is that, in this case, the dependence of $\langle j_\theta \rangle_{ss}$ on $u$ is qualitatively changed from the overdamped model as follows: 
$|\langle j_\theta \rangle_{ss} |$ in the overdamped model has a monotonic dependence on $u$, whereas $|\langle j_\theta \rangle_{ss} |$ with finite inertia is maximized at a specific value of $u$ and becomes smaller as $u$ approaches $k$, as shown in Fig.~\ref{fig3}(b).
For $m=0$, how strongly the two different heat baths are coupled only determines the magnitude of the rotational motion, and thus the monotonic dependence on $u$ is natural. However, when we consider the inertial term, $I_\theta$ is more significant at high $u$ because of the large radial variance $\langle x_1^2 + x_2^2 \rangle_{ss}$, and this makes the rotational motion difficult to perform.
As the combined effects of the two coupled heat baths and $I_\theta$, $\langle j_\theta \rangle_{ss}$ of the inertial model shows a non-monotonic behavior with $u$ in contrast to the overdamped model.

%%%%%%%%%%%%% figure 4 %%%%%%%%%%%%%%
\begin{figure}[t]
    \includegraphics[width=\columnwidth]{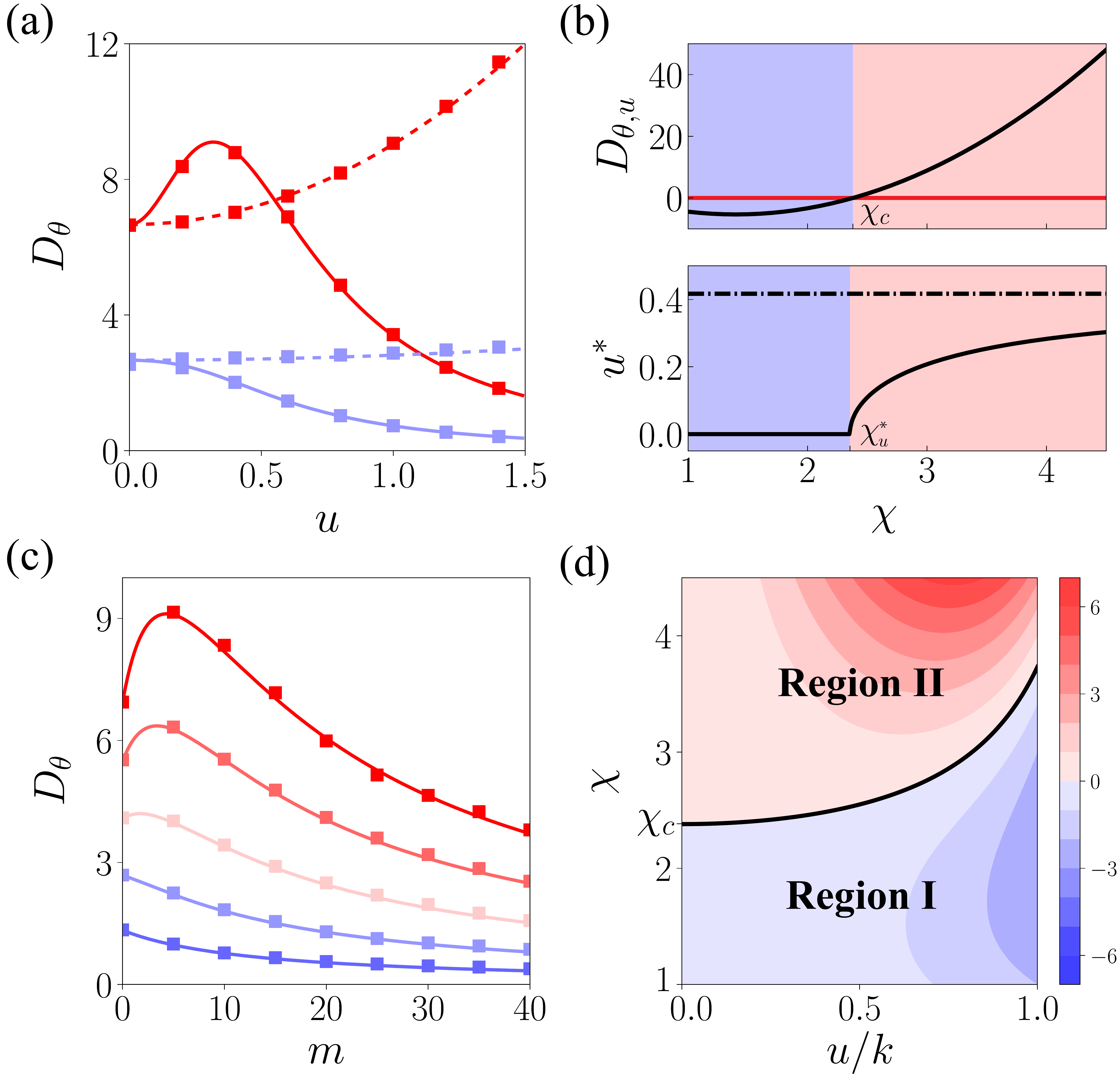}
    \vskip -0.1in
    \caption{ (a) Fluctuation of specific angular momentum $D_\theta$ with $m=5$ (solid) and $D_{\theta, 0}$ (dashed) as functions of $u$. $T_1=5$ (red upper line) and $2$ (blue lower line). (b) Top, bottom: $D_{\theta, u}$ and $u^*$ as a function of $\chi \equiv T_1/T_2$ with $m=5$, respectively. The dashed dotted line in the bottom panel indicates the limit of $u^*$.
    (c) $D_\theta$ as a function of mass $m$ with $u=1/3$ and $T_1 = 1$--$5$ from blue (lower) to red (upper). (d) Contour plot of $D_{\theta, m}$ for $0 < u/k < 1$ and $\chi > 1$. Region \RNum{1} (\RNum{2}) indicates where $D_\theta$ is maximized at $m=0$ (the non-zero $m^*$).
    In the figure, lines and squares represent analytical and Langevin simulation results, respectively. 
    The other parameters are fixed as $T_2=1$, $k=3/2$, and $\gamma = 1$.
    }\label{fig4}
\end{figure}
%%%%%%%%%%%%% figure 4 %%%%%%%%%%%%%%

%%%%%%%%%%%%% figure 5 %%%%%%%%%%%%%%
\begin{figure*}[t]
    \includegraphics[width=\linewidth]{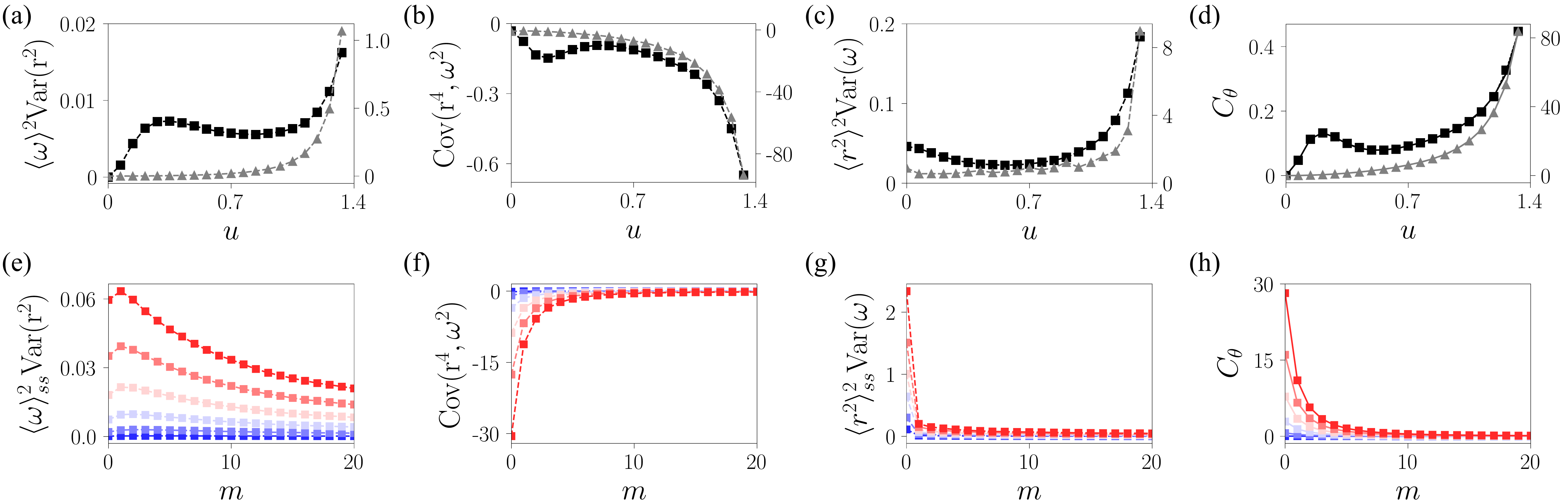}
    \vskip -0.1in
    \caption{Langevin simulation results of four terms, $\langle \omega \rangle^2_{ss} \rm{Var}(r^2)$, $\rm{Cov}(r^4, \omega^2)$, $\langle r^2 \rangle^2_{ss} \rm{Var}(\omega)$, and $C_\theta$, with $\Delta t = 10^{-3}$ and $t=10^3$. (a--d) The four terms as a function of $u$ with $T_1 = 6$ and $m=0$ (gray triangles) or $m=10$ (black squares). Right (left) tick marks indicate the value for $m=0$ ($m=10$). (e--h) The four terms as a function of $m$ with $u=1/2$ and $T_1 = 2$--$12$ from blue (lower) to red (upper). Dashed (solid) lines are simulation (analytical) results and eye guides. The other parameters are fixed as $k=3/2$, $T_2 = 1$, and $\gamma = 1$.}\label{fig5}
\end{figure*}
%%%%%%%%%%%%% figure 7 %%%%%%%%%%%%%%
As the next step to grasp how inertia affects the stochastic dynamics of rotational motion, we consider the fluctuation of $j_\theta(t)$, which is defined as
\begin{equation}\label{eq:Def_SAM-fluct}
\begin{gathered}
    D_\theta \equiv \lim_{t \to \infty} \frac{t}{2} \left( \langle j_\theta (t)^2 \rangle - \langle j_\theta(t) \rangle^2 \right).
\end{gathered}
\end{equation}
The fluctuation with finite inertia, denoted by $D_\theta$, can be explicitly derived as
\begin{equation}\label{eq:InertialSAM_Fluctuation}
\begin{gathered}
    D_\theta = D^{th}_{\theta} + D^{neq}_{\theta},
\end{gathered}
\end{equation}
where
\begin{equation}\label{eq:InertialSAM_Fluctuation2}
\begin{gathered}
    D_{\theta}^{th} = \frac{2\gamma T_1 T_2}{k\gamma^2 + u^2m}, \\
    D_{\theta}^{neq} = \frac{\gamma u^2 (\gamma^4 + u^2m^2 + 5k\gamma^2m)}{2(k\gamma^2 + u^2m)^3}(T_1 - T_2)^2.
\end{gathered}
\end{equation}
To simplify the expressions, $D_{\theta, 0}$ denotes the fluctuation in the overdamped model, which is given by
\begin{equation}\label{eq:Du0}
\begin{aligned}
    D_{\theta, 0} = \frac{2 T_1 T_2}{k\gamma} + \frac{u^2 (T_1 - T_2)^2}{2k^3 \gamma}.
\end{aligned}
\end{equation}
The detailed method is written in Appendix~\ref{sec:Appendix_energy}. We divided $D_\theta$ into two terms, $D_{\theta}^{th}$ and $D_{\theta}^{neq}$. Here, $D_{\theta}^{th}$ is strictly positive in any condition, but $D_{\theta}^{neq}$ appears only in a nonequilibrium state ($T_1 \neq T_2$ and $u \neq 0$). 

As can be seen in Fig.~\ref{fig4}(a), $D_\theta$ has an obviously different curve from $D_{\theta, 0}$ along $u$;
while $D_{\theta, 0}$ monotonically increases with $u$, $D_\theta$ decreases as $u$ approaches $k$. 
This difference comes from the fact that $I_\theta$ becomes large as $u$ approaches $k$ so that the variance of the rotational motion should be small at large $u$.
The remarkable point is that $D_\theta$ is maximized at a non-zero $u^*$, which means the stability of rotation is minimized at this specific $u$.

Furthermore, in a certain range of $u$, $D_\theta$ exceeds $D_{\theta, 0}$ under a specific condition, that is $\chi > \chi_{c}$, where $\chi \equiv T_2/T_1$ ($>1$) and the critical temperature ratio is denoted by $\chi_{c}$.
To obtain the analytical expression of $\chi_{c}$, we expand $D_\theta$ as
\begin{equation}
\begin{aligned}
    D_\theta = D_{\theta, 0} + D_{\theta, u} u^2 +\mathcal{O}(u^4),
\end{aligned}
\label{eq:Du_series}
\end{equation}
where
\begin{equation}
\begin{aligned}
    D_{\theta, u} &= \frac{5(T_1 - T_2)^2 - 4 T_1 T_2}{2k^2 \gamma^3}m.
\end{aligned}
\label{eq:dD/du}
\end{equation}
When $D_{\theta, u}$ is positive, $D_\theta$ increases more rapidly than $D_{\theta, 0}$ near $u=0$. Thus, $\chi_{c}$ is obtained as 
\begin{equation}\label{eq:CritTemp_u}
\begin{aligned}
    \chi_{c} = \frac{7+2\sqrt{6}}{5}.
\end{aligned}
\end{equation}
Figure~\ref{fig4}(b) plots $D_{\theta, u}$ (top) and $u^*$ (bottom) as a function of $\chi$.
Note that the minimum value of $\chi$ for non-zero $u^*$ is denoted by $\chi_u^*$, which is less than $\chi_c$. Actually, $\chi_c$ is the upper bound of $\chi_u^*$, where
\begin{equation}
\begin{aligned}
    \chi_u^* = 1 + \frac{2 \left( 1 + \sqrt{6+\gamma^2/km}\right)}{5+\gamma^2/km}.
\end{aligned}
\end{equation}
Here, $\chi_u^*$ has a value in the range $(1, \chi_c)$ according to the value of $\gamma^2/km$.
Thus, $D_\theta$ is maximized at the non-zero $u^*$ and exceeds $D_{\theta, 0}$ for $\chi > \chi_c$ at the same time. The dashed dotted line in the bottom panel of Fig.~\ref{fig4}(b) indicates the limit of $u^*$ given by
\begin{equation}
\begin{aligned}
    u^* < \frac{\gamma}{m}\sqrt{\sqrt{\gamma^4 + 9k\gamma^2 m + 21k^2 m^2}- 4km - \gamma^2}.
\end{aligned}
\label{eq:limit_of_u*}
\end{equation}
If the limit from Eq.~\eqref{eq:limit_of_u*} is larger than $k$, $u^* < k$ due to the existence condition for steady state (see Appendix~\ref{sec:AppendixA}).

A similar behavior is also observed along $m$, as can be seen in Fig.~\ref{fig4}(c). $D_\theta$ is maximized not at $m=0$ but at an optimal mass $m^*$ under a specific condition, which means the stability of rotation can be minimized at this specific mass. To obtain the condition for the non-zero $m^*$, we expand $D_\theta$ near $m=0$ as
\begin{equation}\label{eq:Dm_series}
\begin{aligned}
    D_\theta &= D_{\theta, 0} + D_{\theta, m}m + \mathcal{O}(m^2),
\end{aligned}
\end{equation}
where
\begin{equation}\label{eq:dD/dm}
\begin{aligned}
    D_{\theta, m} &= \frac{\left( (5k^2 - 3u^2)(T_1-T_2)^2 - 4k^2 T_1 T_2 \right) u^2}{2k^4 \gamma^3}.
\end{aligned}
\end{equation}
Since $D_{\theta, m}$ should be positive for the non-zero $m^*$, $D_\theta$ can be separated into two regions, as illustrated in Fig.~\ref{fig4}(d), where $D_{\theta, m}$ has a negative (positive) value in region \RNum{1} (II) and thus $D_\theta$ decreases (increases) as $m$ increases. 
Intriguingly, Eq.~\eqref{eq:Dm_series} matches Eq.~\eqref{eq:Du_series} for $u^2 \ll k^2$, i.e. $D_{\theta, m}m \simeq D_{\theta, u}u^2$. This correspondence leads to similar non-monotonic behaviors of $D_\theta$ along $u$ and $m$ as well as reveals the condition $\chi_c$ at the boundary of regions \RNum{1} and \RNum{2}, as indicated in Fig.~\ref{fig4}(d).

While we analytically showed that $D_\theta$ has increasing parts with $u$ and $m$ under specific conditions, it is not easy to clarify what induces these behaviors through the expression of $D_\theta$ alone.
To explain the origin of the non-monotonic curves, we divide $D_\theta$ into four terms as
\begin{equation}
\begin{aligned}
    D_\theta &\simeq \lim_{t \to \infty} \frac{t}{2} ( \rm{Cov}(r^4, \omega^2) + \langle \omega \rangle^2_{ss} \rm{Var}(r^2) \\
    &+ \langle r^2 \rangle^2_{ss} \rm{Var}(\omega)
    + C_\theta ),
\label{eq:AppD_Dtheta}
\end{aligned}
\end{equation}
where the cycling frequency is defined as $\omega \equiv (\bm{x} \times \bm{v})/r^2$, $\rm{Cov}(r^4, \omega^2) \equiv \langle r^4 \omega^2 \rangle_{ss} - \langle r^4 \rangle_{ss} \langle \omega^2 \rangle_{ss}$, and  $C_\theta \equiv \langle r^2 \rangle^2_{ss} \langle \omega \rangle^2_{ss} - \langle r^2 \omega \rangle^2_{ss}$. Here, we neglect $\rm{Var}(r^2)\rm{Var}(\omega)$. Because only $C_\theta$ is calculable and determined independent of time, we measure the other terms by Langevin simulations with time step $\Delta t = 10^{-3}$ and total simulation time $t=10^3$. Figure~\ref{fig5} depicts the Langevin simulation results for $\langle \omega \rangle^2_{ss} \rm{Var}(r^2)$, $\rm{Cov}(r^4, \omega^2)$, $\langle r^2 \rangle^2_{ss} \rm{Var}(\omega)$, and $C_\theta$. 
Although the magnitudes of $\rm{Cov}(r^4, \omega^2)$ and $C_\theta$ are much larger than the others, we should consider all terms since the true behavior of $D_\theta$ will not be revealed if we neglect any of the terms.

Let us firstly discuss $D_\theta$ in terms of $u$. The divergence of $\langle r^2 \rangle_{ss}$ in the limit of $u \rightarrow k$ results in large magnitudes of all terms as $u$ approaches $k$, as shown in Fig.~\ref{fig5}(a--d). This increase at large $u$ is a common property between the inertial and overdamped models. The main discrepancy between the two models, though, is that there are local optimum points at specific $u$ caused by $\langle \omega \rangle_{ss}$. Since the inertia of the particle shifts the peak of $\langle \omega \rangle_{ss}$ to lower $u$ (see Fig.~\ref{fig7} in Appendix~\ref{sec:Appendix_Torque&SAM}), we find curved shapes of $\langle \omega \rangle^2_{ss} \rm{Var}(r^2)$, $\rm{Cov}(r^4, \omega^2)$, and $C_\theta$.
We note that $\langle r^2 \rangle_{ss}$ is proportional to $T_1 + T_2$ whereas $\langle \omega \rangle_{ss}$ increases with $|T_1 - T_2|$. Therefore, when $|T_1 - T_2|$ is not sufficiently large, the curved shapes do not appear and the crossover between $\langle r^2 \rangle_{ss}$ and $\langle \omega \rangle_{ss}$ leads to the non-trivial behavior of $D_\theta$ with $u$, as indicated in Fig.~\ref{fig4}(a).

It is more complicated to describe the behavior of $D_\theta$ in terms of $m$ since $D_\theta$ is determined by the joint effects of several factors. As $\langle \omega \rangle_{ss}$ and $\rm{Var}(\omega)$ become smaller with $m$, $\langle r^2 \rangle^2_{ss}\rm{Var}(\omega)$ and $C_\theta$ monotonically decrease with $m$, as shown in Fig.~\ref{fig5}(g,h). 
In the case of $\langle \omega \rangle^2_{ss} \rm{Var}(r^2)$, $I_\theta$ has a role to resist changes in rotational motion, and this induces an increase of radial variance $\rm{Var}(r^2)$ with $m$~\cite{scholz2018inertial}. Thus, $\langle \omega \rangle^2_{ss}\rm{Var}(r^2)$ has a curved shape as illustrated in Fig.~\ref{fig5}(e). 
$\rm{Cov}(r^4, \omega^2)$ has a negative value due to the elliptical shape of the rotational motion (Figs.~\ref{fig1} and~\ref{fig2}), but because the elliptical shape becomes circular and the magnitude of $\langle j_\theta \rangle_{ss}$ decreases with increasing $m$, the magnitude of $\rm{Cov}(r^4, \omega^2)$ approaches zero (i.e. increases), as shown in Fig.~\ref{fig5}(f).
Therefore, we can conclude that the increases of $\rm{Var}(r^2)$ and $\rm{Cov}(r^4, \omega^2)$ along with the decrease of $\langle \omega \rangle_{ss}$ together produce the non-trivial curves of $D_\theta$ with $m$.

\section{Stochastic Energetics: relation to specific angular momentum}
\label{sec:4}

To this point we have focused on how inertia affects the system dynamics of a Brownian gyrator, i.e., a tilted PDF and a specific angular momentum. In the present section, we analytically calculate energetic quantities and show that the system dynamics is highly related to the system energetics. Using this relation, it is possible to infer the behavior of energetic quantities through accessible variables in an underdamped regime.
In models with multiple heat baths, such as a Brownian gyrator, estimations of the energetic quantities using the stochastic area tensor~\cite{gonzalez2019experimental} and the cycling frequency~\cite{mura2018nonequilibrium} have been reported, but they are restricted to the overdamped regime.
Therefore, we extend the previous studies to the underdamped regime and additionally calculate the fluctuations of the energetic quantities in this section.
%%%%%%%%%%%%% figure 6 %%%%%%%%%%%%%%
\begin{figure}[t]
    \includegraphics[width=\columnwidth]{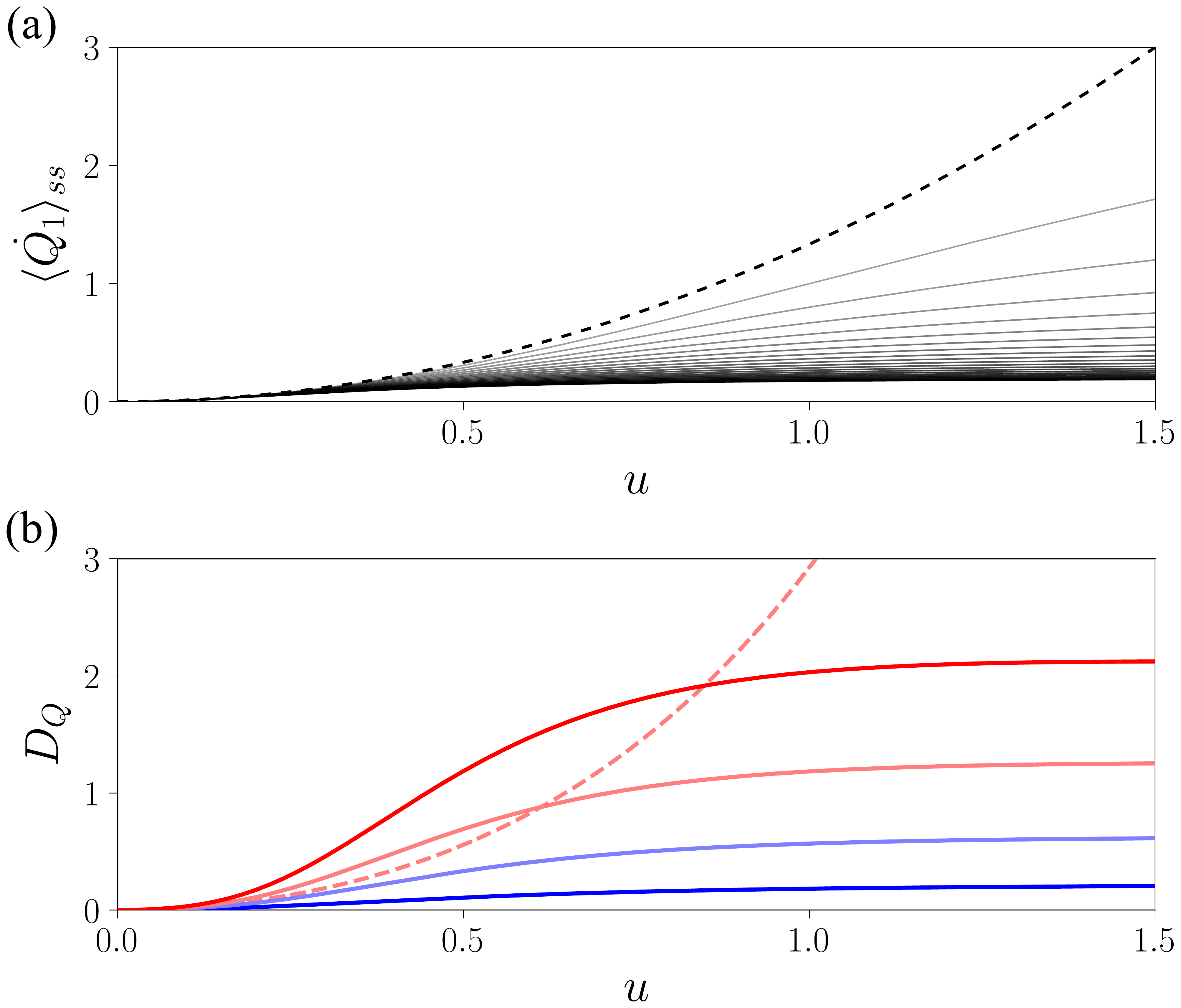}
    \vskip -0.1in
    \caption{ (a) Mean heat current $\langle \dot{Q}_1 \rangle_{ss}$ as a function of $u$. $\langle \dot{Q}_1 \rangle_{ss}$ is plotted for $m = 0$--$10$ with $T_1 = 5$. The solid gray lines darken as $m$ increases. (b) Fluctuation of heat $D_Q$ as a function of $u$. $D_Q$ is drawn with $T_1 = 2$, $4$, $6$, and $8$ from blue (lower) to red (upper) with $m=5$. Solid and dashed lines represent the inertial and overdamped models, respectively. The other parameters are fixed as $T_2=1$, $k=3/2$, and $\gamma = 1$.
    }\label{fig6}
\end{figure}
%%%%%%%%%%%%% figure 6 %%%%%%%%%%%%%%

To calculate the mean current of the absorbed heat, let $\dbar Q_i$ denote the absorbed heat from the environment along $x_i$ over time interval $dt$. Then the absorbed heat is written as
\begin{equation}\label{eq:OverHeat}
\begin{aligned}
    \dbar Q_i = \left( -\gamma v_i(t) + \xi_i(t) \right) \circ dx_i(t),
\end{aligned}
\end{equation}
where $dx_i (t)$ is the evolution of $x_i (t)$ over $dt$ and $\circ$ denotes the Stratonovich product~\cite{sekimoto2010stochastic}. Using the Langevin equation with finite inertia [Eq.~\eqref{eq:InertialLangevinEq}], $\dbar Q_i$ can be expressed in different ways such as $\dbar Q_i = \left( m \dot{v}_i(t) +  \nabla_{x_i}U(\bm{x}) \right) \circ dx_i(t)$. Thus, we obtain the absorbed heat $\dbar Q_{i}$ as follows~\cite{chun2015hidden}:
\begin{equation}\label{eq:InertialHeat2}
\begin{aligned}
    \dbar Q_1 &= d \left( \frac{1}{2}mv_1^2 + \frac{1}{2} k x_1^2 + \frac{1}{2}u x_1 x_2 \right) \\& - \frac{u}{2}\left( x_1 v_2 - x_2 v_1 \right) dt, \\
    \dbar Q_2 &= d \left( \frac{1}{2}mv_2^2 +\frac{1}{2} k x_2^2 + \frac{1}{2}u x_1 x_2 \right) \\& + \frac{u}{2} \left( x_1 v_2 - x_2 v_1 \right) dt.
\end{aligned}
\end{equation}
Since the change of internal energy defined as $dE = \dbar Q_1 + \dbar Q_2 = d \left( m\bm{v}^2/2 + U(\bm{x}) \right)$ cannot affect the steady-state average, the mean heat currents in the steady state are given by
\begin{equation}\label{eq:InertialHeat3}
\begin{aligned}
    \langle \dot{Q}_2 \rangle_{ss} &= \frac{u}{2} \langle x_1 v_2 - x_2 v_1 \rangle_{ss} \\
    &= \frac{u}{2} \langle j_\theta \rangle_{ss} = \frac{\gamma u^2 (T_2 - T_1)}{2(k\gamma^2 + u^2 m)},
\end{aligned}
\end{equation}
where $\langle \dot{Q}_2 \rangle_{ss}=-\langle \dot{Q}_1 \rangle_{ss}$, which reflects the energy conservation law; more specifically, the energy absorbed from the hot bath is equally dissipated to the cold bath. In our system, $\langle \dot{Q}_1 \rangle_{ss}$ is positive due to $T_1 > T_2$. 
Additionally, the exchanged heat between the two heat baths is proportional to $\langle j_\theta \rangle_{ss}$ and vanishes at $u=0$. 
These tendencies imply that an anisotropic potential connects the two different heat baths and converts the rotational motion of the particle to heat that is exchanged between the baths.
This exchanged heat increases the total entropy production, which is defined by the sum of the system and medium entropy productions. The total production rate in the steady state $\langle \dot{S} \rangle_{ss}$ is obtained by
\begin{equation}\label{eq:InertialEP}
\begin{aligned}
    \langle \dot{S} \rangle_{ss} &= -\frac{\langle \dot{Q}_1 \rangle_{ss}}{T_1} - \frac{\langle \dot{Q}_2 \rangle_{ss}}{T_2} \\
    &= \left( \frac{1}{T_1} - \frac{1}{T_2} \right) \frac{u}{2} \langle j_\theta \rangle_{ss}
    = \frac{\gamma u^2 (T_2 - T_1)^2}{2T_1 T_2 \left(k\gamma^2 + u^2 m\right)}.
\end{aligned}
\end{equation}
Because the system entropy production rate vanishes in steady state, $\langle \dot{S} \rangle_{ss}$ coincides with the medium entropy production rate, $\langle \dot{S}_m \rangle_{ss}$.
Total entropy production can be used to quantify the violation of detailed balance and has a positive value in a NESS. Here, $\langle \dot{S} \rangle_{ss} \geq 0$ and the equality is satisfied when $u=0$ or $T_1 = T_2$, that is the condition for the NESS of our system.
As shown in Eqs.~\eqref{eq:InertialHeat3} and \eqref{eq:InertialEP}, $\langle \dot{Q}_i \rangle_{ss}$ and $\langle \dot{S} \rangle_{ss}$ are proportional to $\langle j_\theta \rangle_{ss}$. Note that we only need the variances of the positional trajectory of the particle to measure $\langle j_\theta \rangle_{ss}$. Therefore, the system energetics can be easily obtained from only the positional trajectories without observing the velocity fields.

Next, let us evaluate the fluctuations of the energetic quantities, which is necessary to understand the stochastic properties of the energetics or to design a reliable heat engine.
Fluctuation of absorbed heat $D_{Q_i}$ is defined as
\begin{equation}\label{eq:Def_Heat-fluct}
\begin{gathered}
    D_{Q} = D_{Q_1} = D_{Q_2} = \lim_{t \to \infty} \frac{t}{2} \left( \langle \dot{Q}_{1(2)}(t)^2 \rangle - \langle \dot{Q}_{1(2)}(t) \rangle^2 \right).
\end{gathered}
\end{equation}
As indicated in the equation, $D_{Q_1}$ and $D_{Q_2}$ are the same since the absorbed heats are symmetrically coupled by $u$.
The relation between $D_Q$ and $D_\theta$ can be obtained as
\begin{equation}\label{eq:InertialHeatFlcut}
\begin{gathered}
    D_{Q} = \frac{u^2}{4} D_{\theta}.
\end{gathered}
\end{equation}
The detailed calculation is written in Appendix~\ref{sec:Appendix_energy}. In the same way, fluctuation of the medium entropy production $D_{S_m}$ is given by
\begin{equation}\label{eq:InertialEPFlcut}
\begin{gathered}
    D_{S_m} = \left(\frac{1}{T_1} - \frac{1}{T_2} \right)^2\frac{u^2}{4} D_{\theta}.
\end{gathered}
\end{equation}
Thus, $D_Q$ and $D_{S_m}$ are also proportional to $D_\theta$, similar to the result with the mean currents. 

According to Eqs.~\eqref{eq:InertialHeat3}, \eqref{eq:InertialEP}, \eqref{eq:InertialHeatFlcut}, and \eqref{eq:InertialEPFlcut}, we know that the mean current and the fluctuations of heat and medium entropy production have the same curves as $\langle j_\theta \rangle_{ss}$ and $D_\theta$ as a function of $m$: $\langle \dot{Q}_1 \rangle_{ss}$ and $\langle \dot{S} \rangle_{ss}$ become smaller with increasing $m$, and $D_Q$ and $D_{S_m}$ are maximized at $m^*$ and exceed the overdamped ones in region \RNum{2} as depicted in Fig.~\ref{fig4}(d). Despite these similarities not only with $m$ but also other parameters, only the dependence on $u$ differs, as shown in Fig.~\ref{fig6}(a) and (b). $\langle \dot{Q}_1 \rangle_{ss}$ and $D_{Q}$ are not typically maximized at moderate $u$, while their magnitudes increase with $u$ in contrast to $D_\theta$. Nonetheless, some important features still remain. First, $| \langle \dot{Q}_1 \rangle_{ss} |$ is sufficiently smaller than the value from $m=0$, $\partial D_Q/\partial u$ decreases as $u$ increases, and most interestingly, $D_Q$ exceeds the fluctuation of the overdamped model at the same condition as $D_\theta$, which is $\chi > \chi_c$. 

Although we do not consider external force in this paper to simplify the situation, we can extend our result to a system with applied external force. To extract work from the Brownian gyrator, we must exert an external force in the opposite direction of the rotation of the particle.
When we choose a linear non-conservative force $\bm{F}_{ext}(\bm{x}(t)) = K(x_2, -x_1)^T$ with a constant $K$~\cite{park2016efficiency, pietzonka2018universal, chun2019effect}, the work current $\dot{W}(t)$ is written as
\begin{equation}\label{eq:Work&SAM}
\begin{aligned}
    \dot{W}(t) \equiv \bm{F}_{ext}(\bm{x}(t)) \cdot \bm{\nu}(t) = -K j_\theta (t).
\end{aligned}
\end{equation}
Thus, investigating $j_\theta(t)$ is equivalent to investigating the applied work current.

\section{Conclusions}
\label{sec:5}
In our work, we have examined how Langevin dynamics describes the rotating particle of the Brownian gyrator by explicitly considering the inertia of the particle, in contrast to its description with Brownian dynamics. 
Several NESS features of the Brownian gyrator (such as a tilted PDF, rotational motion, and entropy production by the heat current between the two heat baths) distinguish the inertial model from the overdamped model, i.e. Langevin dynamics from Brownian dynamics.

From the analytic solution of the Fokker--Planck equation and the simulation of each model, we have shown that the inertia plays an important role in resisting the breakdown of the detailed balance and reducing the nonequilibrium effects. 
For instance, in the inertial model, the distortion and tilt of the positional PDF decrease and vanish with increasing mass, starting from the case of the overdamped model. The mean of specific angular momentum $j_\theta$, selected as the measure of the rotational motion and proportional to the systematic torque, also shows similar behavior. 

The most salient feature of our Langevin dynamics description is the non-monotonic behavior of the measure for the rotational motion, which cannot be found through Brownian dynamics.
In the inertial model, the mean of $j_\theta$ has a non-monotonic behavior along $u$. This is because the anisotropy $u$ initiates the rotational motion while the averaged moment of inertia, increasing with $u$, has a contrary role in resisting the rotational motion. 
Next, the fluctuation of $j_\theta$ has non-monotonic behaviors along $u$ and $m$. Intriguingly, we have found that the fluctuation of $j_\theta$ with inertia is larger than the value of the overdamped model in some specific conditions.
These non-monotonic behaviors of the rotation-related quantity appear only in the inertial model. It will be interesting work, therefore, to check whether other nonequilibrium systems have similar non-monotonic features or not.

Considering this intrinsic difference between the two descriptions, i.e. inertial and overdamped models, even in a long time limit, it is necessary to choose Langevin dynamics rather than Brownian dynamics except under the special condition of negligible mass compared to friction. Brownian dynamics may be inadequate to study the Brownian gyrator in experiment, where the particle is not so tiny that it is controllable and observable.

For the successful experimental observation of the inertial effects including the non-monotonic behaviors, the two terms of the denominator of Eq.~\eqref{eq:InertialSAM} should be comparable so that $u^2m/k\gamma^2$ is on the order of $\sim \mathcal{O}(1)$. 
Considering recent experimental achievements~\cite{volpe2006torque, argun2017experimental, leonardo2007parametric}, this condition is actually difficult to attain in a typical liquid environment due to high viscosity. Instead, one may realize this condition in a low-density environment. 
For instance, the condition is satisfied if we consider an optically trapped particle in a gas with optical trapping stiffness $k \sim 1 {\rm pN/\mu m}$~\cite{argun2017experimental}, viscosity $\eta \sim 1$-$10 {\rm \mu Pa \cdot s}$ ($\gamma \sim 10^{-8}$--$10^{-7} {\rm g/s}$)~\cite{leonardo2007parametric}, and the mass of the particle in the range from $10^{-13} \rm{g}$ to $10^{-11} \rm{g}$, comparable to a polystyrene or silica particle with a diameter of $1\ {\rm \mu m}$. 
Then we could see the interesting non-monotonic behaviors by adjusting the anisotropy $u$ (with the same order as $k$) and the mass $m$ in the possible range via optical tweezers. 
For related experiments in a liquid environment, it will be necessary to strengthen the optical trapping stiffness $k$ or $u$, which might be challenging. 

% For the successful observation of the inertial effects including the non-monotonic behaviors in an experiment, the two terms of the denominator of Eq.~\eqref{eq:InertialSAM} should be comparable so that $u^2m/k\gamma^2$ is the order of $\sim \mathcal{O}(1)$. 
% On account of the recent experimental achievements, this condition is actually difficult to be attained in a typical environment using a liquid due to its high viscosity. Instead, one may accomplish this condition in a low-density environment. 
% For instance, the condition is satisfied if we consider an optically trapped particle in a gas with the optical trapping stiffness $k \sim 1 {\rm pN/\mu m}$~\cite{argun2017experimental}, the viscosity $\eta \sim 1$-$10 {\rm \mu Pa \cdot s}$ ($\gamma \sim 10^{-8}$-$10^{-7} {\rm g/s}$)~\cite{leonardo2007parametric} and the mass of a particle in a range from $10^{-13} \rm{g}$ to $10^{-11} \rm{g}$, comparable to a polystyrene or silica particle with diameter $1 {\rm \mu m}$. 
% Then we could see the interesting non-monotonic behaviors by adjusting the anisotropy $u$ (with the same order as $k$) and the mass $m$ in the possible range controlled by optical tweezers. 
% If one insists on an experiment in a liquid environment, it will be necessary to strengthen the optical trapping stiffness $k$ or $u$, which might be challenging.

Finally, we have clarified how the system energetics is associated with the system dynamics in the underdamped regime. We have shown that the mean current and fluctuation of heat currents and entropy production could be inferred using the dynamic characteristics of the system. For example, by observing the tilted angle of the positional PDF or the rotational motion of the particle, it is possible to infer how much energy is exchanged or how much total entropy production increases.

We expect our results to be helpful in studying the dynamic properties of various systems with motion influenced by inertia, such as insects, microflyers, and other mesoscale organisms~\cite{klotsa2019above}. 
Moreover, our results will provide practical ways to investigate various biological systems from an energetic perspective or to quantify the violation of detailed balance. 
In many experimental cases, detailed information is frequently unknown, and it is difficult to estimate the energetics directly. The dynamic characteristics can be adequate measures in such systems for estimating the energetics or inferring their behaviors, as also claimed in Refs.~\cite{mura2018nonequilibrium, gonzalez2019experimental}.
Particularly, $j_\theta$ will be a useful tool for measuring the energetics of tractable nonequilibrium systems because $j_\theta$ is directly associated with them and can be evaluated using experimentally accessible quantities and positional trajectories.

As future work, various models in NESS such as the $N$ bead-spring model or self-propelled particles can be considered to study the inertial effects and the relation between the dynamics and the energetics of the system.

\begin{acknowledgments}
This study was supported by the Basic Science Research Program through the National Research Foundation of Korea (NRF Grant No. 2017R1A2B3006930, 2020R1F1A1076311).

\end{acknowledgments}

\appendix

\section{Steady-state probability density}
\label{sec:AppendixA}
To obtain the steady-state PDF $p(\bm{x}, \bm{v})$, let $\bm{z} \equiv (x_1, x_2, v_1, v_2)^T$ be a state vector. Then we can rewrite Eq.~\eqref{eq:InertialLangevinEq} as
\begin{equation}\label{eq:reLangevin}
    \dot{\bm{z}}(t) = -\mathsf{F} \cdot \bm{z}(t) + \bm{\eta}(t),
\end{equation}
where $\mathsf{F} = \begin{psmallmatrix} \mathsf{0} & -\mathsf{I} \\ -\mathsf{A}/m & \mathsf{\Gamma}/m \end{psmallmatrix}$, $\mathsf{0}$ ($\mathsf{I}$) is a 2 x 2 null (identity) matrix, and $\mathsf{A}$ and $\mathsf{\Gamma}$ are drift and friction matrices, respectively.
$\bm{\eta}$ is a Gaussian white noise satisfying $\langle \eta_i (t) \rangle = 0$ and $\langle \bm{\eta}(t) \cdot \bm{\eta}(t')^T \rangle = 2 \delta(t-t') \mathsf{D}_{\bm{z}}$ where $\mathsf{D}_{\bm{z}} = \begin{psmallmatrix} \mathsf{0} & \mathsf{0} \\ \mathsf{0} & \mathsf{D}\end{psmallmatrix}$ and $\mathsf{D}$ is a diffusion matrix. 
In our system, $\mathsf{A} \equiv -\begin{psmallmatrix} k & u \\ u & k \end{psmallmatrix}$, and $\mathsf{\Gamma} \equiv \begin{psmallmatrix} \gamma & 0 \\ 0 & \gamma \end{psmallmatrix}$ and $\mathsf{D} \equiv \frac{\gamma}{m^2}\begin{psmallmatrix} T_1 & 0 \\ 0 & T_2 \end{psmallmatrix}$ as mentioned in Sec.~\ref{sec:3}.

Since our system assumes an Ornstein--Uhlenbeck process, the steady-state PDF takes a Gaussian form as $p(\bm{z}) \propto \exp{[-(1/2)\bm{z}^T \cdot \mathsf{C}^{-1} \cdot \bm{z}]}$ where $\mathsf{C}$ is the covariance matrix of the state vector $\bm{z}$ in the steady state defined as $\langle \bm{z}\bm{z}^T \rangle_{ss}$. The covariance matrix $\mathsf{C}$ is given by
\begin{equation}\label{eq:kernel}
    \mathsf{C} = \mathsf{F}^{-1}(\mathsf{D}_{\bm{z}} + \mathsf{Q}),
\end{equation}
where $\mathsf{Q}$ is an antisymmetric 4 x 4 matrix that can be uniquely determined by
\begin{equation}\label{eq:Qeq}
    \mathsf{F} \mathsf{Q} + \mathsf{Q} \mathsf{F}^T = \mathsf{F} \mathsf{D}_{\bm{z}} -\mathsf{D}_{\bm{z}} \mathsf{F}^T .
\end{equation}
Here, $\mathsf{Q}$ is non-zero in the NESS, which implies the violation of the detailed balance~\cite{kwon2005structure, kwon2011nonequilibrium}. Solving Eq.~\eqref{eq:Qeq} and inserting the result into Eq.~\eqref{eq:kernel}, we obtain
\begin{widetext}
\begin{equation}\label{eq:4x4CovMatrix}
    \mathsf{C} = 
    \begin{psmallmatrix} \frac{2 k^2 \gamma^2 T_1 + u^2 (k m (T_1 + T_2)+ \gamma^2 (T_2 - T_1))}{2 (k^2-u^2) (k \gamma^2+u^2 m)} & -\frac{u
   (T_1+T_2)}{2 (k^2-u^2)} & 0 & -\frac{u \gamma (T_1-T_2)}{2 (k \gamma^2+u^2 m)} \\
 -\frac{u (T_1+T_2)}{2 (k^2-u^2)} & \frac{2 k^2 \gamma^2 T_2+u^2 (k m (T_1+T_2)+\gamma^2
   (T_1-T_2))}{2 (k^2-u^2) (k \gamma^2+u^2 m)} & \frac{u \gamma (T_1-T_2)}{2 (k \gamma^2+u^2 m)} & 0 \\
 0 & \frac{u \gamma (T_1-T_2)}{2 (k \gamma^2+u^2 m)} & \frac{2 k \gamma^2 T_1+u^2 m (T_1+T_2)}{2 m (k \gamma^2+u^2 m)}
   & 0 \\
 -\frac{u \gamma (T_1-T_2)}{2 (k \gamma^2+u^2 m)} & 0 & 0 & \frac{2 k \gamma^2 T_2+u^2 m (T_1+T_2)}{2 m (k \gamma^2+u^2
   m)} \end{psmallmatrix}.
\end{equation}
\end{widetext}
For the existence of a steady state, $\mathsf{F}$ should be positive-definite. This condition provides the existence condition for steady state $u^2 < k^2$. Otherwise, the particle diverges from the potential.

\section{Relation between the systematic torque and the specific angular momentum}
\label{sec:Appendix_Torque&SAM}

Here we derive the relation between the systematic torque and $j_\theta(t)$ in Eq.~\eqref{eq:Torque&SAM} in the underdamped regime. The systematic torque $M$ that the particle exerts on the potential $U(\bm{x})$ can be expressed as
\begin{equation} \label{eq:AppC_torque}
\begin{aligned}
    M &\equiv \langle \bm{x} \times \nabla_{\bm{x}}U(\bm{x})\rangle_{ss} \\
    &= \int d\bm{x} \int d\bm{v} \;\;\langle \bm{x} \times \nabla_{\bm{x}}U(\bm{x}) | \bm{x}, \bm{v} \rangle_{ss} p(\bm{x}, \bm{v}) \\
    &= - \int d\bm{x} \;\; \bm{x} \times \mathsf{A}\cdot\bm{x} p(\bm{x}),
\end{aligned}
\end{equation}
where $\bm{x} \times \mathsf{A}\cdot\bm{x} = x_1 (\mathsf{A}\cdot\bm{x})_2 - x_2 (\mathsf{A}\cdot\bm{x})_1$.
Solving the Lyapunov equation $\mathsf{F}\mathsf{C} + \mathsf{C}\mathsf{F}^T = 2\mathsf{D}_{\bm{z}}$, we can obtain the following equations:
\begin{align} \label{eq:AppC_Lyapunov1}
    \mathsf{C}_{\bm{x}\bm{v}}^T + \mathsf{C}_{\bm{x}\bm{v}} = 0, \\
    \label{eq:AppC_Lyapunov2}
    \mathsf{C}_{\bm{v}\bm{v}} + \frac{1}{m}\mathsf{A}\mathsf{C}_{\bm{x}\bm{x}} - \frac{1}{m}\mathsf{\Gamma}\mathsf{C}_{\bm{x}\bm{v}}^T= 0,
\end{align}
and
\begin{align}
\frac{1}{m} \left( \mathsf{A}\mathsf{C}_{\bm{x}\bm{v}} + \mathsf{C}_{\bm{x}\bm{v}}^T \mathsf{A} \right) = \frac{1}{m} \left( \mathsf{\Gamma}\mathsf{C}_{\bm{v}\bm{v}} + \mathsf{C}_{\bm{v}\bm{v}} \mathsf{\Gamma} \right) - 2\mathsf{D},
\end{align}
where the covariance matrix $\mathsf{C}$ is denoted by
\begin{equation} \label{eq:AppC_C}
    \mathsf{C} \equiv \begin{pmatrix} \mathsf{C}_{\bm{x}\bm{x}} & \mathsf{C}_{\bm{x}\bm{v}} \\ \mathsf{C}_{\bm{x}\bm{v}}^T & \mathsf{C}_{\bm{v}\bm{v}} \end{pmatrix}.
\end{equation}
Using Eqs.~\eqref{eq:AppC_Lyapunov1} and \eqref{eq:AppC_Lyapunov2}, we can express the drift matrix $\mathsf{A}$ by the covariance matrix, that is,
\begin{equation} \label{eq:AppC_A}
    \mathsf{A} = -m\mathsf{C}_{\bm{v}\bm{v}}\mathsf{C}_{\bm{x}\bm{x}}^{-1} + \gamma \mathsf{C}_{\bm{x}\bm{v}}^T\mathsf{C}_{\bm{x}\bm{x}}^{-1}.
\end{equation}
Inserting Eq.~\eqref{eq:AppC_A} into Eq.~\eqref{eq:AppC_torque}, the systematic torque is written as
\begin{equation} \label{eq:AppC_torque2}
\begin{aligned}
    M &= -\gamma \int d\bm{x} \; \bm{x} \times \mathsf{C}_{\bm{x}\bm{v}}^T\mathsf{C}_{\bm{x}\bm{x}}^{-1} \cdot\bm{x} p(\bm{x}) \\
    &+ m \int d\bm{x} \; \bm{x} \times \mathsf{C}_{\bm{v}\bm{v}}\mathsf{C}_{\bm{x}\bm{x}}^{-1}\cdot\bm{x} p(\bm{x}).
\end{aligned}
\end{equation}
The first term on the right-hand side can be rearranged into a function of $\langle j_{\theta} \rangle_{ss}$ as follows:
\begin{equation} \label{eq:AppC_fst}
\begin{aligned}
    &-\gamma \int d\bm{x} \; \bm{x} \times \mathsf{C}_{\bm{x}\bm{v}}^T\mathsf{C}_{\bm{x}\bm{x}}^{-1} \cdot \bm{x} p(\bm{x}) \\
    &= -\gamma \int d\bm{x} \; \bm{x} \times \langle \bm{v} | \bm{x} \rangle p(\bm{x})
    = -\gamma \langle j_\theta \rangle_{ss}.
\end{aligned}
\end{equation}
In the second row of Eq.~\eqref{eq:AppC_fst}, we use the local averaged velocity $\langle \bm{v} | \bm{x} \rangle = \mathsf{C}_{\bm{x}\bm{v}}^T\mathsf{C}_{\bm{x} \cdot \bm{x}}^{-1} \cdot \bm{x}$~\cite{hogg2005introduction}.
To evaluate the second term on the right-hand side of Eq.~\eqref{eq:AppC_torque2}, let us consider a linear transformation $\bm{x'} = \mathsf{B} \cdot \bm{x}$ where $\mathsf{B}$ is a non-singular matrix. Then, the covariance matrix in the transformed coordinates $\mathsf{C'}_{\bm{x}\bm{x}}$ can be given by $\mathsf{C'}_{\bm{x}\bm{x}} = \mathsf{B}\mathsf{C}_{\bm{x}\bm{x}}\mathsf{B}^T$~\cite{weiss2003coordinate}. If we consider the specific coordinates that satisfy $\mathsf{C'}_{\bm{x}\bm{x}} = \mathsf{I}$, the second term on the right-hand side of Eq.~\eqref{eq:AppC_torque2} vanishes:
\begin{equation} \label{eq:AppC_scd}
\begin{aligned}
    &\int d\bm{x} \; \bm{x} \times \mathsf{C}_{\bm{v}\bm{v}}\mathsf{C}_{\bm{x}\bm{x}}^{-1}\cdot \bm{x} p(\bm{x}) \\
    &= \int d\bm{x} \; \begin{pmatrix} -x_2 & x_1 \end{pmatrix} \mathsf{C}_{\bm{v}\bm{v}}\mathsf{C}_{\bm{x}\bm{x}}^{-1} \begin{pmatrix} x_1 \\ x_2 \end{pmatrix} p(\bm{x}) \\
    &= \frac{1}{\det{\mathsf{B}}} \int d\bm{x'} \; \begin{pmatrix} -x'_2 & x'_1 \end{pmatrix} \mathsf{C'}_{\bm{v}\bm{v}} \begin{pmatrix} x'_1 \\ x'_2 \end{pmatrix} p(\bm{x'}) = 0,
\end{aligned}
\end{equation}
since $\mathsf{C'}_{\bm{v}\bm{v}}$ is a symmetric matrix. Consequently, we obtain the relation as 
\begin{equation} \label{eq:AppC_torque_SAM}
    M = -\gamma \langle j_\theta \rangle_{ss}.
\end{equation}

\begin{figure}[t]
    \includegraphics[width=\columnwidth]{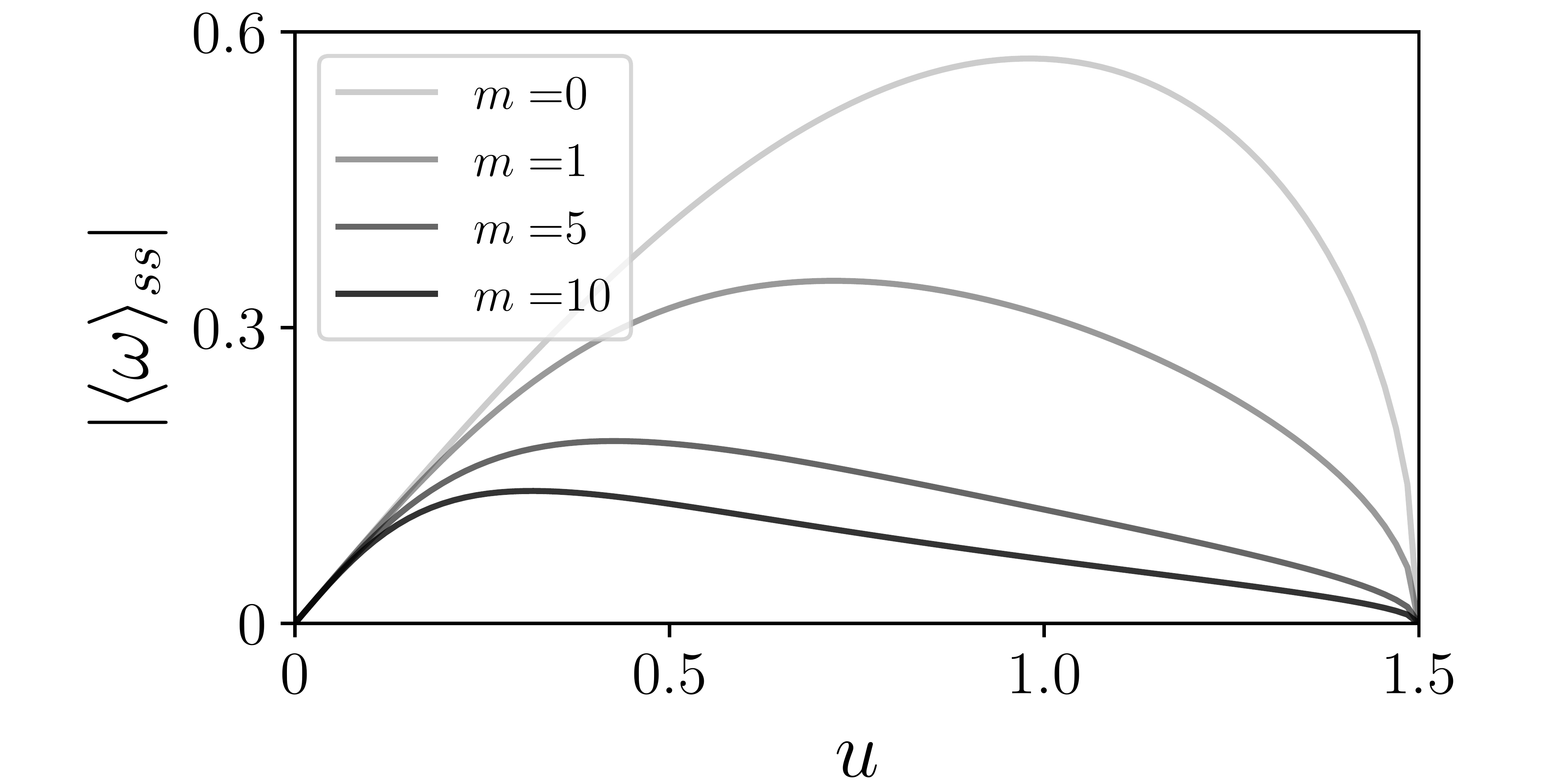}
    \vskip -0.1in
    \caption{ Magnitude of averaged cycling frequency $\langle \omega \rangle_{ss}$ as a function of $u$. $|\langle \omega \rangle_{ss}|$ is plotted for $m=0$, $1$, $5$, and $10$. The other parameters are $T_1 = 5$, $T_2=1$, $k=3/2$, and $\gamma =1$.
    }\label{fig7}
\end{figure}

It is also possible to calculate the cycling frequency by using the relation in the same way as in Ref.~\cite{mura2018nonequilibrium, Gradziuk2019Scaling}. We can express the conditional average $\langle \bm{v}|\bm{x} \rangle$ in another way as $\langle \bm{v}|\bm{x} \rangle = \mathsf{\Omega} \cdot \bm{x}$, where $\mathsf{\Omega}$ is a matrix of frequencies defined as $\mathsf{\Omega} \equiv \mathsf{C}_{\bm{x}\bm{v}}^T\mathsf{C}_{\bm{x}\bm{x}}^{-1}$~\cite{weiss2003coordinate}. In the transformed coordinates, the matrix of frequencies is given by
\begin{equation} \label{eq:AppC_MatFrequency}
    \mathsf{\Omega'} = \begin{pmatrix} 0 & -\alpha \\ \alpha & 0 \end{pmatrix},
\end{equation}
since $\mathsf{\Omega'}$ is skew-symmetric. Thus, the eigenvalues of $\mathsf{\Omega}$ can be obtained by $\lambda = \pm i\alpha$ where $\alpha$ is a real number. Then, $\langle j_\theta \rangle_{ss}$ can be calculated as
\begin{equation} \label{eq:AppC_omega}
\begin{aligned}
     \langle j_\theta \rangle_{ss} &= \int d\bm{x} \; \bm{x} \times \mathsf{\Omega} \cdot \bm{x} p(\bm{x}) \\
     &= \frac{1}{\det{\mathsf{B}}}\int d\bm{x'} \; \bm{x'} \times \mathsf{\Omega'} \cdot \bm{x'} p(\bm{x})\\
     &= \frac{\alpha}{\det{\mathsf{B}}} \langle {x'}_1^2 + {x'}_2^2 \rangle_{ss} = 2 \frac{\alpha}{\det{\mathsf{B}}}.
\end{aligned}
\end{equation}
Using the fact that $\alpha$ is equal to the cycling frequency $\langle \omega \rangle_{ss}$ defined as $\omega(t) = (\bm{x}(t) \times \bm{v}(t))/r(t)^2$ ~\cite{Gradziuk2019Scaling}, the cycling frequency can be calculated as
\begin{equation} \label{eq:AppC_omega2}
\begin{gathered}
     \langle \omega \rangle_{ss} = \frac{\langle j_\theta \rangle_{ss}}{2\sqrt{\det{\mathsf{C}_{\bm{x}\bm{x}}}}} \\
    = \frac{\gamma u (T_2 - T_1)\sqrt{k^2 - u^2}}{\sqrt{\left( u^2 m + k \gamma^2 \right)^2 (T_1 + T_2)^2 - \gamma^4 (k^2 - u^2)(T_1 - T_2)^2 }}.
\end{gathered}
\end{equation}
Here, we used $\det{\mathsf{B}} = \sqrt{\det{\mathsf{C'}_{\bm{x}\bm{x}}}/\det{\mathsf{C}_{\bm{x}\bm{x}}}}$ to derive Eq.~\eqref{eq:AppC_omega2}. As illustrated in Fig.~\ref{fig7}, the magnitude of $\langle \omega \rangle_{ss}$ decreases with $m$ and the peak of $\langle \omega \rangle_{ss}$ is shifted to lower $u$ as $m$ increases because $I_\theta$ becomes larger as $m$ and $u$ increase. 

%%%%%%%%%%%%% figure 7 %%%%%%%%%%%%%%
%%%%%%%%%%%%% figure 7 %%%%%%%%%%%%%%

\section{Mean and fluctuation of energetic quantities}
\label{sec:Appendix_energy}

In this section, we present how to calculate the fluctuations of $j_\theta$, absorbed heat, and medium entropy production using the scaled cumulant generating function (SCGF). We follow the same procedure as in Refs.~\cite{touchette2018introduction, pietzonka2018universal, chun2019effect}. Because the work current $\dot{W}(t)$ is proportional to $j_\theta(t)$, we calculate the mean current and fluctuation of work instead of calculating the values of $j_\theta(t)$ directly.  

When we apply an external force defined by $\bm{F}_{ext}(\bm{x}(t)) = K(x_2, -x_1)^T$, the extracted work, absorbed heat, and medium entropy production over time $\tau$ are given by
\begin{equation}\label{eq:AppC_DynamicaObservables}
\begin{gathered}
    W(\tau) = \int_{0}^{\tau} \mathsf{W} \cdot \bm{z}(t) \circ d\bm{z}(t), \\
    Q_{1 (2)}(\tau) = \int_{0}^{\tau} \mathsf{Q}_{1(2)} \cdot \bm{z}(t) \circ d\bm{z}(t),
\end{gathered}
\end{equation}
and
\begin{equation}\label{eq:AppC_DynamicaObservables2}
\begin{gathered}
    S_m(\tau) = -\int_{0}^{\tau} \left(\frac{\mathsf{Q_1}}{T_1} + \frac{\mathsf{Q_2}}{T_2}\right)\cdot\bm{z}(t) \circ d\bm{z}(t)
\end{gathered}
\end{equation}
where the matrices $\mathsf{W}$ and $\mathsf{Q}_{1 (2)}$ are defined as
\begin{align}
    \label{eq:AppC_SCGFmatrixW}
    \mathsf{W} = 
        \begin{pmatrix} 0 & K & 0 & 0 \\ -K & 0 & 0 & 0 \\ 0 & 0 & 0 & 0 \\ 0 & 0 & 0 & 0 \end{pmatrix}, \\
    \label{eq:AppC_SCGFmatrixQ1}
    \mathsf{Q}_1 = 
        \begin{pmatrix} k & u - K & 0 & 0 \\ 0 & 0 & 0 & 0 \\ 0 & 0 & m & 0 \\ 0 & 0 & 0 & 0 \end{pmatrix} ,
\end{align}
and
\begin{equation}\label{eq:AppC_SCGFmatrixQ2}
\begin{gathered}
    \mathsf{Q}_2 = 
        \begin{pmatrix} 0 & 0 & 0 & 0 \\ u+K & k & 0 & 0 \\ 0 & 0 & 0 & 0 \\ 0 & 0 & 0 & m \end{pmatrix}.
\end{gathered}
\end{equation}
They are dynamical observables due to the dependence on the trajectory over time $\tau$. Thus, the SCGF of work is written as
\begin{equation}\label{eq:AppC_SCGF}
\begin{aligned}
    \lambda (h) &= \lim_{\tau \to \infty} \frac{1}{\tau} \ln{\langle \exp{(h W(\tau))} \rangle} \\
                &=  \langle \dot{W} \rangle_{ss} h + D_W h^2 + \mathcal{O}(h^3),
\end{aligned}
\end{equation}
with a real valued $h$. Here, the mean current and fluctuation of work are denoted as $\langle \dot{W} \rangle_{ss}$ and $D_W$, respectively. It is known that $\lambda (h)$ is the largest eigenvalue of the tilted operator given by
\begin{equation} \label{eq:AppC_tiltedFP}
\begin{aligned}
    \mathcal{L}^\dagger
    &= -\bm{z}^T \cdot \mathsf{F}^{T} \cdot 
    \left( \nabla_{\bm{z}} + h\mathsf{W} \cdot \bm{z} \right)
    \\&+ \left( \nabla_{\bm{z}} + h\mathsf{W}\cdot\bm{z} \right)^T
    \cdot \mathsf{D}_{\bm{z}} \cdot \left( \nabla_{\bm{z}} + h\mathsf{W} \cdot \bm{z}\right),
\end{aligned}
\end{equation}
with $\nabla_{\bm{z}} \equiv \partial/\partial\bm{z}$. Equation~\eqref{eq:AppC_SCGF} can be derived from the fact that the quantity $\langle \exp{(h W(t))}\rangle$ has a semi-group property and is governed by tilted operator [Eq.~\eqref{eq:AppC_tiltedFP}]. This is the so-called Feynman--Kac formula~\cite{touchette2018introduction}. Assuming that the left eigenfunction $g(\bm{z}, h)$ of $\lambda(h)$ is Gaussian as $g(\bm{z}, h) = \exp{\left( -(1/2)\bm{z}^T \cdot \mathsf{G}(h) \cdot \bm{z} \right)}$ with a symmetric matrix $\mathsf{G}(h)$, the SCGF can be expressed as
\begin{equation}\label{eq:AppC_EigenEq}
\begin{aligned}
    \lambda (h) &= (\mathcal{L}^{\dagger}(h) g(\bm{z}, h))/g(\bm{z}, h) \\
            &= \tr{\{\mathsf{D}_{\bm{z}} [h\mathsf{W}-\mathsf{G}]\}} + \bm{z}^T \cdot \mathsf{F}^T (\mathsf{G} - h\mathsf{W}) \cdot \bm{z} \\
            &+ \bm{z}^T \cdot (h\mathsf{W}- \mathsf{G})^T \mathsf{D}_{\bm{z}} (h\mathsf{W}-\mathsf{G})  \cdot \bm{z}.
\end{aligned}
\end{equation}
Comparing the coefficients, we can obtain
\begin{equation} \label{eq:AppC_SCGF2}
    \lambda(h) = \tr{\{\mathsf{D}_{\bm{z}} [h\mathsf{W}-\mathsf{G}(h)]\}},
\end{equation}
and
\begin{equation} \label{eq:AppC_G}
    \mathsf{F}^T (h\mathsf{W}-\mathsf{G})+(h\mathsf{W}-\mathsf{G})^T \mathsf{F} = 2(h\mathsf{W}- \mathsf{G})^T \mathsf{D}_{\bm{z}} (h\mathsf{W}-\mathsf{G}).
\end{equation}
To solve Eq.~\eqref{eq:AppC_G}, we expand $\mathsf{G}(h)$ near $h=0$ as
\begin{equation} \label{eq:AppC_G_taylor}
\begin{gathered}
    \mathsf{G}(h) = \mathsf{G}_1 h + \mathsf{G_2}h^2 + \mathcal{O}(h^3).
\end{gathered}
\end{equation}
Here, the constant term of $\mathsf{G}(h)$ is zero since $\lambda(h) = 0$ with $g(\bm{z}, h)=1$~\cite{touchette2018introduction}. 
Inserting Eq.~\eqref{eq:AppC_G_taylor} into Eq.~\eqref{eq:AppC_G} and comparing the coefficients of $h$ and $h^2$, the mean current and fluctuation of work are obtained by
\begin{equation} \label{eq:AppB_W_cur}
    \langle \dot{W} \rangle_{ss} = \tr{\{ \mathsf{D}_{\bm{z}}(\mathsf{W}-\mathsf{G_1})\}} = \tr{\{ \mathsf{F}\mathsf{C}\mathsf{W}_a \}},
\end{equation}
and
\begin{equation} \label{eq:AppB_W_fluct}
    D_W = -\tr{\{ \mathsf{D}_{\bm{z}} \mathsf{G_2} \}} = \tr{\{ \mathsf{F}\mathsf{C}\mathsf{W}_a \mathsf{C} (\mathsf{W}^T - \mathsf{G}_1) \}},
\end{equation}
where $\mathsf{W}_a = \left(\mathsf{W}-\mathsf{W}^T \right)/2$ is a skew-symmetric matrix. We used the Lyapunov equation $\mathsf{F}\mathsf{C} + \mathsf{C}\mathsf{F}^T = 2\mathsf{D}_{\bm{z}}$ to derive Eqs.~\eqref{eq:AppB_W_cur} and \eqref{eq:AppB_W_fluct}.

Since the work done by the external force $\bm{F}_{ext}$ is proportional to $j_\theta(t)$, we can calculate the mean and fluctuation of $j_\theta(t)$ as 
\begin{equation} \label{eq:SAM_current_fluctuation}
    \langle j_\theta \rangle_{ss} \equiv \lim_{K \to 0} \frac{-\langle \dot{W} \rangle_{ss} }{K} \, , \;\;  D_\theta = \lim_{K \to 0} \frac{D_W}{K^2}.
\end{equation}
In the same way, the mean current and fluctuation of the absorbed heat and medium entropy production can also be obtained by using the matrix defined in Eqs.~\eqref{eq:AppC_SCGFmatrixQ1} and \eqref{eq:AppC_SCGFmatrixQ2}.

\bibliography{InertialEffects}

\begin{thebibliography}{69}%
\makeatletter
\providecommand \@ifxundefined [1]{%
 \@ifx{#1\undefined}
}%
\providecommand \@ifnum [1]{%
 \ifnum #1\expandafter \@firstoftwo
 \else \expandafter \@secondoftwo
 \fi
}%
\providecommand \@ifx [1]{%
 \ifx #1\expandafter \@firstoftwo
 \else \expandafter \@secondoftwo
 \fi
}%
\providecommand \natexlab [1]{#1}%
\providecommand \enquote  [1]{``#1''}%
\providecommand \bibnamefont  [1]{#1}%
\providecommand \bibfnamefont [1]{#1}%
\providecommand \citenamefont [1]{#1}%
\providecommand \href@noop [0]{\@secondoftwo}%
\providecommand \href [0]{\begingroup \@sanitize@url \@href}%
\providecommand \@href[1]{\@@startlink{#1}\@@href}%
\providecommand \@@href[1]{\endgroup#1\@@endlink}%
\providecommand \@sanitize@url [0]{\catcode `\\12\catcode `\$12\catcode
  `\&12\catcode `\#12\catcode `\^12\catcode `\_12\catcode `\%12\relax}%
\providecommand \@@startlink[1]{}%
\providecommand \@@endlink[0]{}%
\providecommand \url  [0]{\begingroup\@sanitize@url \@url }%
\providecommand \@url [1]{\endgroup\@href {#1}{\urlprefix }}%
\providecommand \urlprefix  [0]{URL }%
\providecommand \Eprint [0]{\href }%
\providecommand \doibase [0]{https://doi.org/}%
\providecommand \selectlanguage [0]{\@gobble}%
\providecommand \bibinfo  [0]{\@secondoftwo}%
\providecommand \bibfield  [0]{\@secondoftwo}%
\providecommand \translation [1]{[#1]}%
\providecommand \BibitemOpen [0]{}%
\providecommand \bibitemStop [0]{}%
\providecommand \bibitemNoStop [0]{.\EOS\space}%
\providecommand \EOS [0]{\spacefactor3000\relax}%
\providecommand \BibitemShut  [1]{\csname bibitem#1\endcsname}%
\let\auto@bib@innerbib\@empty
%</preamble>
\bibitem [{\citenamefont {Weber}\ \emph {et~al.}(2012)\citenamefont {Weber},
  \citenamefont {Spakowitz},\ and\ \citenamefont
  {Theriot}}]{weber2012nonthermal}%
  \BibitemOpen
  \bibfield  {author} {\bibinfo {author} {\bibfnamefont {S.~C.}\ \bibnamefont
  {Weber}}, \bibinfo {author} {\bibfnamefont {A.~J.}\ \bibnamefont
  {Spakowitz}},\ and\ \bibinfo {author} {\bibfnamefont {J.~A.}\ \bibnamefont
  {Theriot}},\ }\bibfield  {title} {\bibinfo {title} {Nonthermal atp-dependent
  fluctuations contribute to the in vivo motion of chromosomal loci},\ }\href
  {https://doi.org/10.1073/pnas.1119505109} {\bibfield  {journal} {\bibinfo
  {journal} {Proc. Natl. Acad. Sci. U.S.A.}\ }\textbf {\bibinfo {volume}
  {109}},\ \bibinfo {pages} {7338} (\bibinfo {year} {2012})}\BibitemShut
  {NoStop}%
\bibitem [{\citenamefont {Battle}\ \emph {et~al.}(2015)\citenamefont {Battle},
  \citenamefont {Ott}, \citenamefont {Burnette}, \citenamefont
  {Lippincott-Schwartz},\ and\ \citenamefont
  {Schmidt}}]{battle2015intracellular}%
  \BibitemOpen
  \bibfield  {author} {\bibinfo {author} {\bibfnamefont {C.}~\bibnamefont
  {Battle}}, \bibinfo {author} {\bibfnamefont {C.~M.}\ \bibnamefont {Ott}},
  \bibinfo {author} {\bibfnamefont {D.~T.}\ \bibnamefont {Burnette}}, \bibinfo
  {author} {\bibfnamefont {J.}~\bibnamefont {Lippincott-Schwartz}},\ and\
  \bibinfo {author} {\bibfnamefont {C.~F.}\ \bibnamefont {Schmidt}},\
  }\bibfield  {title} {\bibinfo {title} {Intracellular and extracellular forces
  drive primary cilia movement},\ }\href
  {https://doi.org/10.1073/pnas.1421845112} {\bibfield  {journal} {\bibinfo
  {journal} {Proc. Natl. Acad. Sci. U.S.A.}\ }\textbf {\bibinfo {volume}
  {112}},\ \bibinfo {pages} {1410} (\bibinfo {year} {2015})}\BibitemShut
  {NoStop}%
\bibitem [{\citenamefont {Battle}\ \emph {et~al.}(2016)\citenamefont {Battle},
  \citenamefont {Broedersz}, \citenamefont {Fakhri}, \citenamefont {Geyer},
  \citenamefont {Howard}, \citenamefont {Schmidt},\ and\ \citenamefont
  {MacKintosh}}]{battle2016broken}%
  \BibitemOpen
  \bibfield  {author} {\bibinfo {author} {\bibfnamefont {C.}~\bibnamefont
  {Battle}}, \bibinfo {author} {\bibfnamefont {C.~P.}\ \bibnamefont
  {Broedersz}}, \bibinfo {author} {\bibfnamefont {N.}~\bibnamefont {Fakhri}},
  \bibinfo {author} {\bibfnamefont {V.~F.}\ \bibnamefont {Geyer}}, \bibinfo
  {author} {\bibfnamefont {J.}~\bibnamefont {Howard}}, \bibinfo {author}
  {\bibfnamefont {C.~F.}\ \bibnamefont {Schmidt}},\ and\ \bibinfo {author}
  {\bibfnamefont {F.~C.}\ \bibnamefont {MacKintosh}},\ }\bibfield  {title}
  {\bibinfo {title} {Broken detailed balance at mesoscopic scales in active
  biological systems},\ }\href {https://doi.org/10.1126/science.aac8167}
  {\bibfield  {journal} {\bibinfo  {journal} {Science}\ }\textbf {\bibinfo
  {volume} {352}},\ \bibinfo {pages} {604} (\bibinfo {year}
  {2016})}\BibitemShut {NoStop}%
\bibitem [{\citenamefont {Gov}(2004)}]{gov2004membrane}%
  \BibitemOpen
  \bibfield  {author} {\bibinfo {author} {\bibfnamefont {N.}~\bibnamefont
  {Gov}},\ }\bibfield  {title} {\bibinfo {title} {Membrane undulations driven
  by force fluctuations of active proteins},\ }\href
  {https://doi.org/10.1103/PhysRevLett.93.268104} {\bibfield  {journal}
  {\bibinfo  {journal} {Phys. Rev. Lett.}\ }\textbf {\bibinfo {volume} {93}},\
  \bibinfo {pages} {268104} (\bibinfo {year} {2004})}\BibitemShut {NoStop}%
\bibitem [{\citenamefont {Ben-Isaac}\ \emph {et~al.}(2011)\citenamefont
  {Ben-Isaac}, \citenamefont {Park}, \citenamefont {Popescu}, \citenamefont
  {Brown}, \citenamefont {Gov},\ and\ \citenamefont
  {Shokef}}]{ben2011effective}%
  \BibitemOpen
  \bibfield  {author} {\bibinfo {author} {\bibfnamefont {E.}~\bibnamefont
  {Ben-Isaac}}, \bibinfo {author} {\bibfnamefont {Y.}~\bibnamefont {Park}},
  \bibinfo {author} {\bibfnamefont {G.}~\bibnamefont {Popescu}}, \bibinfo
  {author} {\bibfnamefont {F.~L.~H.}\ \bibnamefont {Brown}}, \bibinfo {author}
  {\bibfnamefont {N.~S.}\ \bibnamefont {Gov}},\ and\ \bibinfo {author}
  {\bibfnamefont {Y.}~\bibnamefont {Shokef}},\ }\bibfield  {title} {\bibinfo
  {title} {Effective temperature of red-blood-cell membrane fluctuations},\
  }\href {https://doi.org/10.1103/PhysRevLett.106.238103} {\bibfield  {journal}
  {\bibinfo  {journal} {Phys. Rev. Lett.}\ }\textbf {\bibinfo {volume} {106}},\
  \bibinfo {pages} {238103} (\bibinfo {year} {2011})}\BibitemShut {NoStop}%
\bibitem [{\citenamefont {Mizuno}\ \emph {et~al.}(2007)\citenamefont {Mizuno},
  \citenamefont {Tardin}, \citenamefont {Schmidt},\ and\ \citenamefont
  {MacKintosh}}]{mizuno2007nonequilibrium}%
  \BibitemOpen
  \bibfield  {author} {\bibinfo {author} {\bibfnamefont {D.}~\bibnamefont
  {Mizuno}}, \bibinfo {author} {\bibfnamefont {C.}~\bibnamefont {Tardin}},
  \bibinfo {author} {\bibfnamefont {C.~F.}\ \bibnamefont {Schmidt}},\ and\
  \bibinfo {author} {\bibfnamefont {F.~C.}\ \bibnamefont {MacKintosh}},\
  }\bibfield  {title} {\bibinfo {title} {Nonequilibrium mechanics of active
  cytoskeletal networks},\ }\href {https://doi.org/10.1126/science.1134404}
  {\bibfield  {journal} {\bibinfo  {journal} {Science}\ }\textbf {\bibinfo
  {volume} {315}},\ \bibinfo {pages} {370} (\bibinfo {year}
  {2007})}\BibitemShut {NoStop}%
\bibitem [{\citenamefont {Gladrow}\ \emph {et~al.}(2016)\citenamefont
  {Gladrow}, \citenamefont {Fakhri}, \citenamefont {MacKintosh}, \citenamefont
  {Schmidt},\ and\ \citenamefont {Broedersz}}]{gladrow2016broken}%
  \BibitemOpen
  \bibfield  {author} {\bibinfo {author} {\bibfnamefont {J.}~\bibnamefont
  {Gladrow}}, \bibinfo {author} {\bibfnamefont {N.}~\bibnamefont {Fakhri}},
  \bibinfo {author} {\bibfnamefont {F.~C.}\ \bibnamefont {MacKintosh}},
  \bibinfo {author} {\bibfnamefont {C.~F.}\ \bibnamefont {Schmidt}},\ and\
  \bibinfo {author} {\bibfnamefont {C.~P.}\ \bibnamefont {Broedersz}},\
  }\bibfield  {title} {\bibinfo {title} {Broken detailed balance of filament
  dynamics in active networks},\ }\href
  {https://doi.org/10.1103/PhysRevLett.116.248301} {\bibfield  {journal}
  {\bibinfo  {journal} {Phys. Rev. Lett.}\ }\textbf {\bibinfo {volume} {116}},\
  \bibinfo {pages} {248301} (\bibinfo {year} {2016})}\BibitemShut {NoStop}%
\bibitem [{\citenamefont {Gladrow}\ \emph {et~al.}(2017)\citenamefont
  {Gladrow}, \citenamefont {Broedersz},\ and\ \citenamefont
  {Schmidt}}]{gladrow2017nonequilibrium}%
  \BibitemOpen
  \bibfield  {author} {\bibinfo {author} {\bibfnamefont {J.}~\bibnamefont
  {Gladrow}}, \bibinfo {author} {\bibfnamefont {C.~P.}\ \bibnamefont
  {Broedersz}},\ and\ \bibinfo {author} {\bibfnamefont {C.~F.}\ \bibnamefont
  {Schmidt}},\ }\bibfield  {title} {\bibinfo {title} {Nonequilibrium dynamics
  of probe filaments in actin-myosin networks},\ }\href
  {https://doi.org/10.1103/PhysRevE.96.022408} {\bibfield  {journal} {\bibinfo
  {journal} {Phys. Rev. E}\ }\textbf {\bibinfo {volume} {96}},\ \bibinfo
  {pages} {022408} (\bibinfo {year} {2017})}\BibitemShut {NoStop}%
\bibitem [{\citenamefont {Lander}\ \emph {et~al.}(2012)\citenamefont {Lander},
  \citenamefont {Mehl}, \citenamefont {Blickle}, \citenamefont {Bechinger},\
  and\ \citenamefont {Seifert}}]{lander2012noninvasive}%
  \BibitemOpen
  \bibfield  {author} {\bibinfo {author} {\bibfnamefont {B.}~\bibnamefont
  {Lander}}, \bibinfo {author} {\bibfnamefont {J.}~\bibnamefont {Mehl}},
  \bibinfo {author} {\bibfnamefont {V.}~\bibnamefont {Blickle}}, \bibinfo
  {author} {\bibfnamefont {C.}~\bibnamefont {Bechinger}},\ and\ \bibinfo
  {author} {\bibfnamefont {U.}~\bibnamefont {Seifert}},\ }\bibfield  {title}
  {\bibinfo {title} {Noninvasive measurement of dissipation in colloidal
  systems},\ }\href {https://doi.org/10.1103/PhysRevE.86.030401} {\bibfield
  {journal} {\bibinfo  {journal} {Phys. Rev. E}\ }\textbf {\bibinfo {volume}
  {86}},\ \bibinfo {pages} {030401} (\bibinfo {year} {2012})}\BibitemShut
  {NoStop}%
\bibitem [{\citenamefont {Gnesotto}\ \emph {et~al.}(2018)\citenamefont
  {Gnesotto}, \citenamefont {Mura}, \citenamefont {Gladrow},\ and\
  \citenamefont {Broedersz}}]{gnesotto2018broken}%
  \BibitemOpen
  \bibfield  {author} {\bibinfo {author} {\bibfnamefont {F.}~\bibnamefont
  {Gnesotto}}, \bibinfo {author} {\bibfnamefont {F.}~\bibnamefont {Mura}},
  \bibinfo {author} {\bibfnamefont {J.}~\bibnamefont {Gladrow}},\ and\ \bibinfo
  {author} {\bibfnamefont {C.}~\bibnamefont {Broedersz}},\ }\bibfield  {title}
  {\bibinfo {title} {Broken detailed balance and non-equilibrium dynamics in
  living systems: a review},\ }\href {https://doi.org/10.1088/1361-6633/aab3ed}
  {\bibfield  {journal} {\bibinfo  {journal} {Rep. Prog. Phys.}\ }\textbf
  {\bibinfo {volume} {81}},\ \bibinfo {pages} {066601} (\bibinfo {year}
  {2018})}\BibitemShut {NoStop}%
\bibitem [{\citenamefont {Baiesi}\ \emph {et~al.}(2009)\citenamefont {Baiesi},
  \citenamefont {Maes},\ and\ \citenamefont
  {Wynants}}]{baiesi2009fluctuations}%
  \BibitemOpen
  \bibfield  {author} {\bibinfo {author} {\bibfnamefont {M.}~\bibnamefont
  {Baiesi}}, \bibinfo {author} {\bibfnamefont {C.}~\bibnamefont {Maes}},\ and\
  \bibinfo {author} {\bibfnamefont {B.}~\bibnamefont {Wynants}},\ }\bibfield
  {title} {\bibinfo {title} {Fluctuations and response of nonequilibrium
  states},\ }\href {https://doi.org/10.1103/PhysRevLett.103.010602} {\bibfield
  {journal} {\bibinfo  {journal} {Phys. Rev. Lett.}\ }\textbf {\bibinfo
  {volume} {103}},\ \bibinfo {pages} {010602} (\bibinfo {year}
  {2009})}\BibitemShut {NoStop}%
\bibitem [{\citenamefont {Sharma}\ and\ \citenamefont
  {Brader}(2016)}]{sharma2016communication}%
  \BibitemOpen
  \bibfield  {author} {\bibinfo {author} {\bibfnamefont {A.}~\bibnamefont
  {Sharma}}\ and\ \bibinfo {author} {\bibfnamefont {J.~M.}\ \bibnamefont
  {Brader}},\ }\bibfield  {title} {\bibinfo {title} {Communication: Green-kubo
  approach to the average swim speed in active brownian systems},\ }\href
  {https://doi.org/10.1063/1.4966153} {\bibfield  {journal} {\bibinfo
  {journal} {J. Chem. Phys.}\ }\textbf {\bibinfo {volume} {145}},\ \bibinfo
  {pages} {161101} (\bibinfo {year} {2016})}\BibitemShut {NoStop}%
\bibitem [{\citenamefont {Asheichyk}\ \emph {et~al.}(2019)\citenamefont
  {Asheichyk}, \citenamefont {Solon}, \citenamefont {Rohwer},\ and\
  \citenamefont {Kr{\"u}ger}}]{asheichyk2019response}%
  \BibitemOpen
  \bibfield  {author} {\bibinfo {author} {\bibfnamefont {K.}~\bibnamefont
  {Asheichyk}}, \bibinfo {author} {\bibfnamefont {A.~P.}\ \bibnamefont
  {Solon}}, \bibinfo {author} {\bibfnamefont {C.~M.}\ \bibnamefont {Rohwer}},\
  and\ \bibinfo {author} {\bibfnamefont {M.}~\bibnamefont {Kr{\"u}ger}},\
  }\bibfield  {title} {\bibinfo {title} {Response of active brownian particles
  to shear flow},\ }\href {https://doi.org/10.1063/1.5086495} {\bibfield
  {journal} {\bibinfo  {journal} {J. Chem. Phys.}\ }\textbf {\bibinfo {volume}
  {150}},\ \bibinfo {pages} {144111} (\bibinfo {year} {2019})}\BibitemShut
  {NoStop}%
\bibitem [{\citenamefont {Gallavotti}\ and\ \citenamefont
  {Cohen}(1995)}]{gallavotti1995dynamical}%
  \BibitemOpen
  \bibfield  {author} {\bibinfo {author} {\bibfnamefont {G.}~\bibnamefont
  {Gallavotti}}\ and\ \bibinfo {author} {\bibfnamefont {E.~G.~D.}\ \bibnamefont
  {Cohen}},\ }\bibfield  {title} {\bibinfo {title} {Dynamical ensembles in
  nonequilibrium statistical mechanics},\ }\href
  {https://doi.org/10.1103/PhysRevLett.74.2694} {\bibfield  {journal} {\bibinfo
   {journal} {Phys. Rev. Lett.}\ }\textbf {\bibinfo {volume} {74}},\ \bibinfo
  {pages} {2694} (\bibinfo {year} {1995})}\BibitemShut {NoStop}%
\bibitem [{\citenamefont {Kurchan}(1998)}]{kurchan1998fluctuation}%
  \BibitemOpen
  \bibfield  {author} {\bibinfo {author} {\bibfnamefont {J.}~\bibnamefont
  {Kurchan}},\ }\bibfield  {title} {\bibinfo {title} {Fluctuation theorem for
  stochastic dynamics},\ }\href {https://doi.org/10.1088/0305-4470/31/16/003}
  {\bibfield  {journal} {\bibinfo  {journal} {J. Phys. A.}\ }\textbf {\bibinfo
  {volume} {31}},\ \bibinfo {pages} {3719} (\bibinfo {year}
  {1998})}\BibitemShut {NoStop}%
\bibitem [{\citenamefont {Seifert}(2005)}]{seifert2005entropy}%
  \BibitemOpen
  \bibfield  {author} {\bibinfo {author} {\bibfnamefont {U.}~\bibnamefont
  {Seifert}},\ }\bibfield  {title} {\bibinfo {title} {Entropy production along
  a stochastic trajectory and an integral fluctuation theorem},\ }\href
  {https://doi.org/10.1103/PhysRevLett.95.040602} {\bibfield  {journal}
  {\bibinfo  {journal} {Phys. Rev. Lett.}\ }\textbf {\bibinfo {volume} {95}},\
  \bibinfo {pages} {040602} (\bibinfo {year} {2005})}\BibitemShut {NoStop}%
\bibitem [{\citenamefont {Seifert}(2012)}]{seifert2012stochastic}%
  \BibitemOpen
  \bibfield  {author} {\bibinfo {author} {\bibfnamefont {U.}~\bibnamefont
  {Seifert}},\ }\bibfield  {title} {\bibinfo {title} {Stochastic
  thermodynamics, fluctuation theorems and molecular machines},\ }\href
  {https://doi.org/10.1088/0034-4885/75/12/126001} {\bibfield  {journal}
  {\bibinfo  {journal} {Rep. Prog. Phys.}\ }\textbf {\bibinfo {volume} {75}},\
  \bibinfo {pages} {126001} (\bibinfo {year} {2012})}\BibitemShut {NoStop}%
\bibitem [{\citenamefont {Barato}\ and\ \citenamefont
  {Seifert}(2015)}]{barato2015thermodynamic}%
  \BibitemOpen
  \bibfield  {author} {\bibinfo {author} {\bibfnamefont {A.~C.}\ \bibnamefont
  {Barato}}\ and\ \bibinfo {author} {\bibfnamefont {U.}~\bibnamefont
  {Seifert}},\ }\bibfield  {title} {\bibinfo {title} {Thermodynamic uncertainty
  relation for biomolecular processes},\ }\href
  {https://doi.org/10.1103/PhysRevLett.114.158101} {\bibfield  {journal}
  {\bibinfo  {journal} {Phys. Rev. Lett.}\ }\textbf {\bibinfo {volume} {114}},\
  \bibinfo {pages} {158101} (\bibinfo {year} {2015})}\BibitemShut {NoStop}%
\bibitem [{\citenamefont {Gingrich}\ \emph {et~al.}(2016)\citenamefont
  {Gingrich}, \citenamefont {Horowitz}, \citenamefont {Perunov},\ and\
  \citenamefont {England}}]{gingrich2016dissipation}%
  \BibitemOpen
  \bibfield  {author} {\bibinfo {author} {\bibfnamefont {T.~R.}\ \bibnamefont
  {Gingrich}}, \bibinfo {author} {\bibfnamefont {J.~M.}\ \bibnamefont
  {Horowitz}}, \bibinfo {author} {\bibfnamefont {N.}~\bibnamefont {Perunov}},\
  and\ \bibinfo {author} {\bibfnamefont {J.~L.}\ \bibnamefont {England}},\
  }\bibfield  {title} {\bibinfo {title} {Dissipation bounds all steady-state
  current fluctuations},\ }\href
  {https://doi.org/10.1103/PhysRevLett.116.120601} {\bibfield  {journal}
  {\bibinfo  {journal} {Phys. Rev. Lett.}\ }\textbf {\bibinfo {volume} {116}},\
  \bibinfo {pages} {120601} (\bibinfo {year} {2016})}\BibitemShut {NoStop}%
\bibitem [{\citenamefont {Pietzonka}\ \emph {et~al.}(2017)\citenamefont
  {Pietzonka}, \citenamefont {Ritort},\ and\ \citenamefont
  {Seifert}}]{pietzonka2017finite}%
  \BibitemOpen
  \bibfield  {author} {\bibinfo {author} {\bibfnamefont {P.}~\bibnamefont
  {Pietzonka}}, \bibinfo {author} {\bibfnamefont {F.}~\bibnamefont {Ritort}},\
  and\ \bibinfo {author} {\bibfnamefont {U.}~\bibnamefont {Seifert}},\
  }\bibfield  {title} {\bibinfo {title} {Finite-time generalization of the
  thermodynamic uncertainty relation},\ }\href
  {https://doi.org/10.1103/PhysRevE.96.012101} {\bibfield  {journal} {\bibinfo
  {journal} {Phys. Rev. E}\ }\textbf {\bibinfo {volume} {96}},\ \bibinfo
  {pages} {012101} (\bibinfo {year} {2017})}\BibitemShut {NoStop}%
\bibitem [{\citenamefont {Pietzonka}\ and\ \citenamefont
  {Seifert}(2018)}]{pietzonka2018universal}%
  \BibitemOpen
  \bibfield  {author} {\bibinfo {author} {\bibfnamefont {P.}~\bibnamefont
  {Pietzonka}}\ and\ \bibinfo {author} {\bibfnamefont {U.}~\bibnamefont
  {Seifert}},\ }\bibfield  {title} {\bibinfo {title} {Universal trade-off
  between power, efficiency, and constancy in steady-state heat engines},\
  }\href {https://doi.org/10.1103/PhysRevLett.120.190602} {\bibfield  {journal}
  {\bibinfo  {journal} {Phys. Rev. Lett.}\ }\textbf {\bibinfo {volume} {120}},\
  \bibinfo {pages} {190602} (\bibinfo {year} {2018})}\BibitemShut {NoStop}%
\bibitem [{\citenamefont {Chun}\ \emph {et~al.}(2019)\citenamefont {Chun},
  \citenamefont {Fischer},\ and\ \citenamefont {Seifert}}]{chun2019effect}%
  \BibitemOpen
  \bibfield  {author} {\bibinfo {author} {\bibfnamefont {H.-M.}\ \bibnamefont
  {Chun}}, \bibinfo {author} {\bibfnamefont {L.~P.}\ \bibnamefont {Fischer}},\
  and\ \bibinfo {author} {\bibfnamefont {U.}~\bibnamefont {Seifert}},\
  }\bibfield  {title} {\bibinfo {title} {Effect of a magnetic field on the
  thermodynamic uncertainty relation},\ }\href
  {https://doi.org/10.1103/PhysRevE.99.042128} {\bibfield  {journal} {\bibinfo
  {journal} {Phys. Rev. E}\ }\textbf {\bibinfo {volume} {99}},\ \bibinfo
  {pages} {042128} (\bibinfo {year} {2019})}\BibitemShut {NoStop}%
\bibitem [{\citenamefont {Blum}\ \emph {et~al.}(2006)\citenamefont {Blum},
  \citenamefont {Bruns}, \citenamefont {Rademacher}, \citenamefont {Voss},
  \citenamefont {Willenberg},\ and\ \citenamefont
  {Krause}}]{blum2006measurement}%
  \BibitemOpen
  \bibfield  {author} {\bibinfo {author} {\bibfnamefont {J.}~\bibnamefont
  {Blum}}, \bibinfo {author} {\bibfnamefont {S.}~\bibnamefont {Bruns}},
  \bibinfo {author} {\bibfnamefont {D.}~\bibnamefont {Rademacher}}, \bibinfo
  {author} {\bibfnamefont {A.}~\bibnamefont {Voss}}, \bibinfo {author}
  {\bibfnamefont {B.}~\bibnamefont {Willenberg}},\ and\ \bibinfo {author}
  {\bibfnamefont {M.}~\bibnamefont {Krause}},\ }\bibfield  {title} {\bibinfo
  {title} {Measurement of the translational and rotational brownian motion of
  individual particles in a rarefied gas},\ }\href
  {https://doi.org/10.1103/PhysRevLett.97.230601} {\bibfield  {journal}
  {\bibinfo  {journal} {Phys. Rev. Lett.}\ }\textbf {\bibinfo {volume} {97}},\
  \bibinfo {pages} {230601} (\bibinfo {year} {2006})}\BibitemShut {NoStop}%
\bibitem [{\citenamefont {Li}\ \emph {et~al.}(2010)\citenamefont {Li},
  \citenamefont {Kheifets}, \citenamefont {Medellin},\ and\ \citenamefont
  {Raizen}}]{li2010measurement}%
  \BibitemOpen
  \bibfield  {author} {\bibinfo {author} {\bibfnamefont {T.}~\bibnamefont
  {Li}}, \bibinfo {author} {\bibfnamefont {S.}~\bibnamefont {Kheifets}},
  \bibinfo {author} {\bibfnamefont {D.}~\bibnamefont {Medellin}},\ and\
  \bibinfo {author} {\bibfnamefont {M.~G.}\ \bibnamefont {Raizen}},\ }\bibfield
   {title} {\bibinfo {title} {Measurement of the instantaneous velocity of a
  brownian particle},\ }\href {https://doi.org/10.1126/science.1189403}
  {\bibfield  {journal} {\bibinfo  {journal} {Science}\ }\textbf {\bibinfo
  {volume} {328}},\ \bibinfo {pages} {1673} (\bibinfo {year}
  {2010})}\BibitemShut {NoStop}%
\bibitem [{\citenamefont {Huang}\ \emph {et~al.}(2011)\citenamefont {Huang},
  \citenamefont {Chavez}, \citenamefont {Taute}, \citenamefont {Luki{\'c}},
  \citenamefont {Jeney}, \citenamefont {Raizen},\ and\ \citenamefont
  {Florin}}]{huang2011direct}%
  \BibitemOpen
  \bibfield  {author} {\bibinfo {author} {\bibfnamefont {R.}~\bibnamefont
  {Huang}}, \bibinfo {author} {\bibfnamefont {I.}~\bibnamefont {Chavez}},
  \bibinfo {author} {\bibfnamefont {K.~M.}\ \bibnamefont {Taute}}, \bibinfo
  {author} {\bibfnamefont {B.}~\bibnamefont {Luki{\'c}}}, \bibinfo {author}
  {\bibfnamefont {S.}~\bibnamefont {Jeney}}, \bibinfo {author} {\bibfnamefont
  {M.~G.}\ \bibnamefont {Raizen}},\ and\ \bibinfo {author} {\bibfnamefont
  {E.-L.}\ \bibnamefont {Florin}},\ }\bibfield  {title} {\bibinfo {title}
  {Direct observation of the full transition from ballistic to diffusive
  brownian motion in a liquid},\ }\href {https://doi.org/10.1038/nphys1953}
  {\bibfield  {journal} {\bibinfo  {journal} {Nat. Phys.}\ }\textbf {\bibinfo
  {volume} {7}},\ \bibinfo {pages} {576} (\bibinfo {year} {2011})}\BibitemShut
  {NoStop}%
\bibitem [{\citenamefont {Pusey}(2011)}]{pusey2011brownian}%
  \BibitemOpen
  \bibfield  {author} {\bibinfo {author} {\bibfnamefont {P.~N.}\ \bibnamefont
  {Pusey}},\ }\bibfield  {title} {\bibinfo {title} {Brownian motion goes
  ballistic},\ }\href {https://doi.org/10.1126/science.1192222} {\bibfield
  {journal} {\bibinfo  {journal} {Science}\ }\textbf {\bibinfo {volume}
  {332}},\ \bibinfo {pages} {802} (\bibinfo {year} {2011})}\BibitemShut
  {NoStop}%
\bibitem [{\citenamefont {Attanasi}\ \emph {et~al.}(2014)\citenamefont
  {Attanasi}, \citenamefont {Cavagna}, \citenamefont {Del~Castello},
  \citenamefont {Giardina}, \citenamefont {Grigera}, \citenamefont {Jeli{\'c}},
  \citenamefont {Melillo}, \citenamefont {Parisi}, \citenamefont {Pohl},
  \citenamefont {Shen} \emph {et~al.}}]{attanasi2014information}%
  \BibitemOpen
  \bibfield  {author} {\bibinfo {author} {\bibfnamefont {A.}~\bibnamefont
  {Attanasi}}, \bibinfo {author} {\bibfnamefont {A.}~\bibnamefont {Cavagna}},
  \bibinfo {author} {\bibfnamefont {L.}~\bibnamefont {Del~Castello}}, \bibinfo
  {author} {\bibfnamefont {I.}~\bibnamefont {Giardina}}, \bibinfo {author}
  {\bibfnamefont {T.~S.}\ \bibnamefont {Grigera}}, \bibinfo {author}
  {\bibfnamefont {A.}~\bibnamefont {Jeli{\'c}}}, \bibinfo {author}
  {\bibfnamefont {S.}~\bibnamefont {Melillo}}, \bibinfo {author} {\bibfnamefont
  {L.}~\bibnamefont {Parisi}}, \bibinfo {author} {\bibfnamefont
  {O.}~\bibnamefont {Pohl}}, \bibinfo {author} {\bibfnamefont {E.}~\bibnamefont
  {Shen}}, \emph {et~al.},\ }\bibfield  {title} {\bibinfo {title} {Information
  transfer and behavioural inertia in starling flocks},\ }\href
  {https://doi.org/10.1038/nphys3035} {\bibfield  {journal} {\bibinfo
  {journal} {Nat. Phys.}\ }\textbf {\bibinfo {volume} {10}},\ \bibinfo {pages}
  {691} (\bibinfo {year} {2014})}\BibitemShut {NoStop}%
\bibitem [{\citenamefont {Katz}\ \emph {et~al.}(2011)\citenamefont {Katz},
  \citenamefont {Tunstr{\o}m}, \citenamefont {Ioannou}, \citenamefont {Huepe},\
  and\ \citenamefont {Couzin}}]{katz2011inferring}%
  \BibitemOpen
  \bibfield  {author} {\bibinfo {author} {\bibfnamefont {Y.}~\bibnamefont
  {Katz}}, \bibinfo {author} {\bibfnamefont {K.}~\bibnamefont {Tunstr{\o}m}},
  \bibinfo {author} {\bibfnamefont {C.~C.}\ \bibnamefont {Ioannou}}, \bibinfo
  {author} {\bibfnamefont {C.}~\bibnamefont {Huepe}},\ and\ \bibinfo {author}
  {\bibfnamefont {I.~D.}\ \bibnamefont {Couzin}},\ }\bibfield  {title}
  {\bibinfo {title} {Inferring the structure and dynamics of interactions in
  schooling fish},\ }\href {https://doi.org/10.1073/pnas.1107583108} {\bibfield
   {journal} {\bibinfo  {journal} {Proc. Natl. Acad. Sci. U.S.A.}\ }\textbf
  {\bibinfo {volume} {108}},\ \bibinfo {pages} {18720} (\bibinfo {year}
  {2011})}\BibitemShut {NoStop}%
\bibitem [{\citenamefont {Giomi}\ \emph {et~al.}(2013)\citenamefont {Giomi},
  \citenamefont {Hawley-Weld},\ and\ \citenamefont
  {Mahadevan}}]{giomi2013swarming}%
  \BibitemOpen
  \bibfield  {author} {\bibinfo {author} {\bibfnamefont {L.}~\bibnamefont
  {Giomi}}, \bibinfo {author} {\bibfnamefont {N.}~\bibnamefont {Hawley-Weld}},\
  and\ \bibinfo {author} {\bibfnamefont {L.}~\bibnamefont {Mahadevan}},\
  }\bibfield  {title} {\bibinfo {title} {Swarming, swirling and stasis in
  sequestered bristle-bots},\ }\href
  {https://doi.org/https://doi.org/10.1098/rspa.2012.0637} {\bibfield
  {journal} {\bibinfo  {journal} {Proc. R. Soc. A.}\ }\textbf {\bibinfo
  {volume} {469}},\ \bibinfo {pages} {20120637} (\bibinfo {year}
  {2013})}\BibitemShut {NoStop}%
\bibitem [{\citenamefont {Selmeczi}\ \emph {et~al.}(2005)\citenamefont
  {Selmeczi}, \citenamefont {Mosler}, \citenamefont {Hagedorn}, \citenamefont
  {Larsen},\ and\ \citenamefont {Flyvbjerg}}]{selmeczi2005cell}%
  \BibitemOpen
  \bibfield  {author} {\bibinfo {author} {\bibfnamefont {D.}~\bibnamefont
  {Selmeczi}}, \bibinfo {author} {\bibfnamefont {S.}~\bibnamefont {Mosler}},
  \bibinfo {author} {\bibfnamefont {P.~H.}\ \bibnamefont {Hagedorn}}, \bibinfo
  {author} {\bibfnamefont {N.~B.}\ \bibnamefont {Larsen}},\ and\ \bibinfo
  {author} {\bibfnamefont {H.}~\bibnamefont {Flyvbjerg}},\ }\bibfield  {title}
  {\bibinfo {title} {Cell motility as persistent random motion: theories from
  experiments},\ }\href {https://doi.org/10.1529/biophysj.105.061150}
  {\bibfield  {journal} {\bibinfo  {journal} {Biophys. J.}\ }\textbf {\bibinfo
  {volume} {89}},\ \bibinfo {pages} {912} (\bibinfo {year} {2005})}\BibitemShut
  {NoStop}%
\bibitem [{\citenamefont {Rabault}\ \emph {et~al.}(2019)\citenamefont
  {Rabault}, \citenamefont {Fauli},\ and\ \citenamefont
  {Carlson}}]{rabault2019curving}%
  \BibitemOpen
  \bibfield  {author} {\bibinfo {author} {\bibfnamefont {J.}~\bibnamefont
  {Rabault}}, \bibinfo {author} {\bibfnamefont {R.~A.}\ \bibnamefont {Fauli}},\
  and\ \bibinfo {author} {\bibfnamefont {A.}~\bibnamefont {Carlson}},\
  }\bibfield  {title} {\bibinfo {title} {Curving to fly: Synthetic adaptation
  unveils optimal flight performance of whirling fruits},\ }\href
  {https://doi.org/10.1103/PhysRevLett.122.024501} {\bibfield  {journal}
  {\bibinfo  {journal} {Phys. Rev. Lett.}\ }\textbf {\bibinfo {volume} {122}},\
  \bibinfo {pages} {024501} (\bibinfo {year} {2019})}\BibitemShut {NoStop}%
\bibitem [{\citenamefont {Klotsa}(2019)}]{klotsa2019above}%
  \BibitemOpen
  \bibfield  {author} {\bibinfo {author} {\bibfnamefont {D.}~\bibnamefont
  {Klotsa}},\ }\bibfield  {title} {\bibinfo {title} {As above, so below, and
  also in between: mesoscale active matter in fluids},\ }\href
  {https://doi.org/10.1039/C9SM01019J} {\bibfield  {journal} {\bibinfo
  {journal} {Soft matter}\ }\textbf {\bibinfo {volume} {15}},\ \bibinfo {pages}
  {8946} (\bibinfo {year} {2019})}\BibitemShut {NoStop}%
\bibitem [{\citenamefont {Benjamin}\ and\ \citenamefont
  {Kawai}(2008)}]{benjamin2008inertial}%
  \BibitemOpen
  \bibfield  {author} {\bibinfo {author} {\bibfnamefont {R.}~\bibnamefont
  {Benjamin}}\ and\ \bibinfo {author} {\bibfnamefont {R.}~\bibnamefont
  {Kawai}},\ }\bibfield  {title} {\bibinfo {title} {Inertial effects in
  b\"uttiker-landauer motor and refrigerator at the overdamped limit},\ }\href
  {https://doi.org/10.1103/PhysRevE.77.051132} {\bibfield  {journal} {\bibinfo
  {journal} {Phys. Rev. E}\ }\textbf {\bibinfo {volume} {77}},\ \bibinfo
  {pages} {051132} (\bibinfo {year} {2008})}\BibitemShut {NoStop}%
\bibitem [{\citenamefont {Celani}\ \emph {et~al.}(2012)\citenamefont {Celani},
  \citenamefont {Bo}, \citenamefont {Eichhorn},\ and\ \citenamefont
  {Aurell}}]{celani2012anomalous}%
  \BibitemOpen
  \bibfield  {author} {\bibinfo {author} {\bibfnamefont {A.}~\bibnamefont
  {Celani}}, \bibinfo {author} {\bibfnamefont {S.}~\bibnamefont {Bo}}, \bibinfo
  {author} {\bibfnamefont {R.}~\bibnamefont {Eichhorn}},\ and\ \bibinfo
  {author} {\bibfnamefont {E.}~\bibnamefont {Aurell}},\ }\bibfield  {title}
  {\bibinfo {title} {Anomalous thermodynamics at the microscale},\ }\href
  {https://doi.org/10.1103/PhysRevLett.109.260603} {\bibfield  {journal}
  {\bibinfo  {journal} {Phys. Rev. Lett.}\ }\textbf {\bibinfo {volume} {109}},\
  \bibinfo {pages} {260603} (\bibinfo {year} {2012})}\BibitemShut {NoStop}%
\bibitem [{\citenamefont {Ao}\ \emph {et~al.}(2007)\citenamefont {Ao},
  \citenamefont {Kwon},\ and\ \citenamefont {Qian}}]{ao2007existence}%
  \BibitemOpen
  \bibfield  {author} {\bibinfo {author} {\bibfnamefont {P.}~\bibnamefont
  {Ao}}, \bibinfo {author} {\bibfnamefont {C.}~\bibnamefont {Kwon}},\ and\
  \bibinfo {author} {\bibfnamefont {H.}~\bibnamefont {Qian}},\ }\bibfield
  {title} {\bibinfo {title} {On the existence of potential landscape in the
  evolution of complex systems},\ }\href {https://doi.org/10.1002/cplx.20171}
  {\bibfield  {journal} {\bibinfo  {journal} {Complexity}\ }\textbf {\bibinfo
  {volume} {12}},\ \bibinfo {pages} {19} (\bibinfo {year} {2007})}\BibitemShut
  {NoStop}%
\bibitem [{\citenamefont {Yuan}\ \emph {et~al.}(2017)\citenamefont {Yuan},
  \citenamefont {Tang},\ and\ \citenamefont {Ao}}]{yuan2017sde}%
  \BibitemOpen
  \bibfield  {author} {\bibinfo {author} {\bibfnamefont {R.}~\bibnamefont
  {Yuan}}, \bibinfo {author} {\bibfnamefont {Y.}~\bibnamefont {Tang}},\ and\
  \bibinfo {author} {\bibfnamefont {P.}~\bibnamefont {Ao}},\ }\bibfield
  {title} {\bibinfo {title} {Sde decomposition and a-type stochastic
  interpretation in nonequilibrium processes},\ }\href
  {https://doi.org/10.1007/s11467-017-0718-2} {\bibfield  {journal} {\bibinfo
  {journal} {Front. Phys.}\ }\textbf {\bibinfo {volume} {12}},\ \bibinfo
  {pages} {120201} (\bibinfo {year} {2017})}\BibitemShut {NoStop}%
\bibitem [{\citenamefont {Chun}\ \emph {et~al.}(2018)\citenamefont {Chun},
  \citenamefont {Durang},\ and\ \citenamefont {Noh}}]{Chun2018Emergence}%
  \BibitemOpen
  \bibfield  {author} {\bibinfo {author} {\bibfnamefont {H.-M.}\ \bibnamefont
  {Chun}}, \bibinfo {author} {\bibfnamefont {X.}~\bibnamefont {Durang}},\ and\
  \bibinfo {author} {\bibfnamefont {J.~D.}\ \bibnamefont {Noh}},\ }\bibfield
  {title} {\bibinfo {title} {Emergence of nonwhite noise in langevin dynamics
  with magnetic lorentz force},\ }\href
  {https://doi.org/10.1103/PhysRevE.97.032117} {\bibfield  {journal} {\bibinfo
  {journal} {Phys. Rev. E}\ }\textbf {\bibinfo {volume} {97}},\ \bibinfo
  {pages} {032117} (\bibinfo {year} {2018})}\BibitemShut {NoStop}%
\bibitem [{\citenamefont {Lee}\ and\ \citenamefont
  {Kwon}(2019)}]{Lee2019Nonequilibrium}%
  \BibitemOpen
  \bibfield  {author} {\bibinfo {author} {\bibfnamefont {S.}~\bibnamefont
  {Lee}}\ and\ \bibinfo {author} {\bibfnamefont {C.}~\bibnamefont {Kwon}},\
  }\bibfield  {title} {\bibinfo {title} {Nonequilibrium driven by an external
  torque in the presence of a magnetic field},\ }\href
  {https://doi.org/10.1103/PhysRevE.99.052142} {\bibfield  {journal} {\bibinfo
  {journal} {Phys. Rev. E}\ }\textbf {\bibinfo {volume} {99}},\ \bibinfo
  {pages} {052142} (\bibinfo {year} {2019})}\BibitemShut {NoStop}%
\bibitem [{\citenamefont {Mandal}\ \emph {et~al.}(2019)\citenamefont {Mandal},
  \citenamefont {Liebchen},\ and\ \citenamefont
  {L\"owen}}]{mandal2019motility}%
  \BibitemOpen
  \bibfield  {author} {\bibinfo {author} {\bibfnamefont {S.}~\bibnamefont
  {Mandal}}, \bibinfo {author} {\bibfnamefont {B.}~\bibnamefont {Liebchen}},\
  and\ \bibinfo {author} {\bibfnamefont {H.}~\bibnamefont {L\"owen}},\
  }\bibfield  {title} {\bibinfo {title} {Motility-induced temperature
  difference in coexisting phases},\ }\href
  {https://doi.org/10.1103/PhysRevLett.123.228001} {\bibfield  {journal}
  {\bibinfo  {journal} {Phys. Rev. Lett.}\ }\textbf {\bibinfo {volume} {123}},\
  \bibinfo {pages} {228001} (\bibinfo {year} {2019})}\BibitemShut {NoStop}%
\bibitem [{\citenamefont {Scholz}\ \emph {et~al.}(2018)\citenamefont {Scholz},
  \citenamefont {Jahanshahi}, \citenamefont {Ldov},\ and\ \citenamefont
  {L{\"o}wen}}]{scholz2018inertial}%
  \BibitemOpen
  \bibfield  {author} {\bibinfo {author} {\bibfnamefont {C.}~\bibnamefont
  {Scholz}}, \bibinfo {author} {\bibfnamefont {S.}~\bibnamefont {Jahanshahi}},
  \bibinfo {author} {\bibfnamefont {A.}~\bibnamefont {Ldov}},\ and\ \bibinfo
  {author} {\bibfnamefont {H.}~\bibnamefont {L{\"o}wen}},\ }\bibfield  {title}
  {\bibinfo {title} {Inertial delay of self-propelled particles},\ }\href
  {https://doi.org/10.1038/s41467-018-07596-x} {\bibfield  {journal} {\bibinfo
  {journal} {Nat. Commun.}\ }\textbf {\bibinfo {volume} {9}},\ \bibinfo {pages}
  {5156} (\bibinfo {year} {2018})}\BibitemShut {NoStop}%
\bibitem [{\citenamefont {Dauchot}\ and\ \citenamefont
  {D\'emery}(2019)}]{dauchot2019dynamics}%
  \BibitemOpen
  \bibfield  {author} {\bibinfo {author} {\bibfnamefont {O.}~\bibnamefont
  {Dauchot}}\ and\ \bibinfo {author} {\bibfnamefont {V.}~\bibnamefont
  {D\'emery}},\ }\bibfield  {title} {\bibinfo {title} {Dynamics of a
  self-propelled particle in a harmonic trap},\ }\href
  {https://doi.org/10.1103/PhysRevLett.122.068002} {\bibfield  {journal}
  {\bibinfo  {journal} {Phys. Rev. Lett.}\ }\textbf {\bibinfo {volume} {122}},\
  \bibinfo {pages} {068002} (\bibinfo {year} {2019})}\BibitemShut {NoStop}%
\bibitem [{\citenamefont {Gutierrez-Martinez}\ and\ \citenamefont
  {Sandoval}(2020)}]{gutierrez2020inertial}%
  \BibitemOpen
  \bibfield  {author} {\bibinfo {author} {\bibfnamefont {L.~L.}\ \bibnamefont
  {Gutierrez-Martinez}}\ and\ \bibinfo {author} {\bibfnamefont
  {M.}~\bibnamefont {Sandoval}},\ }\bibfield  {title} {\bibinfo {title}
  {Inertial effects on trapped active matter},\ }\href
  {https://doi.org/https://doi.org/10.1063/5.0011270} {\bibfield  {journal}
  {\bibinfo  {journal} {J. Chem. Phys.}\ }\textbf {\bibinfo {volume} {153}},\
  \bibinfo {pages} {044906} (\bibinfo {year} {2020})}\BibitemShut {NoStop}%
\bibitem [{\citenamefont {L{\"o}wen}(2020)}]{lowen2020inertial}%
  \BibitemOpen
  \bibfield  {author} {\bibinfo {author} {\bibfnamefont {H.}~\bibnamefont
  {L{\"o}wen}},\ }\bibfield  {title} {\bibinfo {title} {Inertial effects of
  self-propelled particles: From active brownian to active langevin motion},\
  }\href {https://doi.org/10.1063/1.5134455} {\bibfield  {journal} {\bibinfo
  {journal} {J. Chem. Phys.}\ }\textbf {\bibinfo {volume} {152}},\ \bibinfo
  {pages} {040901} (\bibinfo {year} {2020})}\BibitemShut {NoStop}%
\bibitem [{\citenamefont {Caprini}\ and\ \citenamefont
  {Marconi}(2021)}]{caprini2020inertial}%
  \BibitemOpen
  \bibfield  {author} {\bibinfo {author} {\bibfnamefont {L.}~\bibnamefont
  {Caprini}}\ and\ \bibinfo {author} {\bibfnamefont {U.~M.~B.}\ \bibnamefont
  {Marconi}},\ }\bibfield  {title} {\bibinfo {title} {Inertial self-propelled
  particles},\ }\href {https://doi.org/10.1063/5.0030940} {\bibfield  {journal}
  {\bibinfo  {journal} {J. Chem. Phys.}\ }\textbf {\bibinfo {volume} {154}},\
  \bibinfo {pages} {024902} (\bibinfo {year} {2021})}\BibitemShut {NoStop}%
\bibitem [{\citenamefont {Filliger}\ and\ \citenamefont
  {Reimann}(2007)}]{filliger2007brownian}%
  \BibitemOpen
  \bibfield  {author} {\bibinfo {author} {\bibfnamefont {R.}~\bibnamefont
  {Filliger}}\ and\ \bibinfo {author} {\bibfnamefont {P.}~\bibnamefont
  {Reimann}},\ }\bibfield  {title} {\bibinfo {title} {Brownian gyrator: A
  minimal heat engine on the nanoscale},\ }\href
  {https://doi.org/10.1103/PhysRevLett.99.230602} {\bibfield  {journal}
  {\bibinfo  {journal} {Phys. Rev. Lett.}\ }\textbf {\bibinfo {volume} {99}},\
  \bibinfo {pages} {230602} (\bibinfo {year} {2007})}\BibitemShut {NoStop}%
\bibitem [{\citenamefont {Mura}\ \emph {et~al.}(2018)\citenamefont {Mura},
  \citenamefont {Gradziuk},\ and\ \citenamefont
  {Broedersz}}]{mura2018nonequilibrium}%
  \BibitemOpen
  \bibfield  {author} {\bibinfo {author} {\bibfnamefont {F.}~\bibnamefont
  {Mura}}, \bibinfo {author} {\bibfnamefont {G.}~\bibnamefont {Gradziuk}},\
  and\ \bibinfo {author} {\bibfnamefont {C.~P.}\ \bibnamefont {Broedersz}},\
  }\bibfield  {title} {\bibinfo {title} {Nonequilibrium scaling behavior in
  driven soft biological assemblies},\ }\href
  {https://doi.org/10.1103/PhysRevLett.121.038002} {\bibfield  {journal}
  {\bibinfo  {journal} {Phys. Rev. Lett.}\ }\textbf {\bibinfo {volume} {121}},\
  \bibinfo {pages} {038002} (\bibinfo {year} {2018})}\BibitemShut {NoStop}%
\bibitem [{\citenamefont {Gradziuk}\ \emph {et~al.}(2019)\citenamefont
  {Gradziuk}, \citenamefont {Mura},\ and\ \citenamefont
  {Broedersz}}]{Gradziuk2019Scaling}%
  \BibitemOpen
  \bibfield  {author} {\bibinfo {author} {\bibfnamefont {G.}~\bibnamefont
  {Gradziuk}}, \bibinfo {author} {\bibfnamefont {F.}~\bibnamefont {Mura}},\
  and\ \bibinfo {author} {\bibfnamefont {C.~P.}\ \bibnamefont {Broedersz}},\
  }\bibfield  {title} {\bibinfo {title} {Scaling behavior of nonequilibrium
  measures in internally driven elastic assemblies},\ }\href
  {https://doi.org/10.1103/PhysRevE.99.052406} {\bibfield  {journal} {\bibinfo
  {journal} {Phys. Rev. E}\ }\textbf {\bibinfo {volume} {99}},\ \bibinfo
  {pages} {052406} (\bibinfo {year} {2019})}\BibitemShut {NoStop}%
\bibitem [{\citenamefont {Ciliberto}\ \emph {et~al.}(2013)\citenamefont
  {Ciliberto}, \citenamefont {Imparato}, \citenamefont {Naert},\ and\
  \citenamefont {Tanase}}]{cilberto2013heat}%
  \BibitemOpen
  \bibfield  {author} {\bibinfo {author} {\bibfnamefont {S.}~\bibnamefont
  {Ciliberto}}, \bibinfo {author} {\bibfnamefont {A.}~\bibnamefont {Imparato}},
  \bibinfo {author} {\bibfnamefont {A.}~\bibnamefont {Naert}},\ and\ \bibinfo
  {author} {\bibfnamefont {M.}~\bibnamefont {Tanase}},\ }\bibfield  {title}
  {\bibinfo {title} {Heat flux and entropy produced by thermal fluctuations},\
  }\href {https://doi.org/10.1103/PhysRevLett.110.180601} {\bibfield  {journal}
  {\bibinfo  {journal} {Phys. Rev. Lett.}\ }\textbf {\bibinfo {volume} {110}},\
  \bibinfo {pages} {180601} (\bibinfo {year} {2013})}\BibitemShut {NoStop}%
\bibitem [{\citenamefont {Ghanta}\ \emph {et~al.}(2017)\citenamefont {Ghanta},
  \citenamefont {Neu},\ and\ \citenamefont
  {Teitsworth}}]{ghanta2017fluctuation}%
  \BibitemOpen
  \bibfield  {author} {\bibinfo {author} {\bibfnamefont {A.}~\bibnamefont
  {Ghanta}}, \bibinfo {author} {\bibfnamefont {J.~C.}\ \bibnamefont {Neu}},\
  and\ \bibinfo {author} {\bibfnamefont {S.}~\bibnamefont {Teitsworth}},\
  }\bibfield  {title} {\bibinfo {title} {Fluctuation loops in noise-driven
  linear dynamical systems},\ }\href
  {https://doi.org/10.1103/PhysRevE.95.032128} {\bibfield  {journal} {\bibinfo
  {journal} {Phys. Rev. E}\ }\textbf {\bibinfo {volume} {95}},\ \bibinfo
  {pages} {032128} (\bibinfo {year} {2017})}\BibitemShut {NoStop}%
\bibitem [{\citenamefont {Argun}\ \emph {et~al.}(2017)\citenamefont {Argun},
  \citenamefont {Soni}, \citenamefont {Dabelow}, \citenamefont {Bo},
  \citenamefont {Pesce}, \citenamefont {Eichhorn},\ and\ \citenamefont
  {Volpe}}]{argun2017experimental}%
  \BibitemOpen
  \bibfield  {author} {\bibinfo {author} {\bibfnamefont {A.}~\bibnamefont
  {Argun}}, \bibinfo {author} {\bibfnamefont {J.}~\bibnamefont {Soni}},
  \bibinfo {author} {\bibfnamefont {L.}~\bibnamefont {Dabelow}}, \bibinfo
  {author} {\bibfnamefont {S.}~\bibnamefont {Bo}}, \bibinfo {author}
  {\bibfnamefont {G.}~\bibnamefont {Pesce}}, \bibinfo {author} {\bibfnamefont
  {R.}~\bibnamefont {Eichhorn}},\ and\ \bibinfo {author} {\bibfnamefont
  {G.}~\bibnamefont {Volpe}},\ }\bibfield  {title} {\bibinfo {title}
  {Experimental realization of a minimal microscopic heat engine},\ }\href
  {https://doi.org/10.1103/PhysRevE.96.052106} {\bibfield  {journal} {\bibinfo
  {journal} {Phys. Rev. E}\ }\textbf {\bibinfo {volume} {96}},\ \bibinfo
  {pages} {052106} (\bibinfo {year} {2017})}\BibitemShut {NoStop}%
\bibitem [{\citenamefont {Chiang}\ \emph {et~al.}(2017)\citenamefont {Chiang},
  \citenamefont {Lee}, \citenamefont {Lai},\ and\ \citenamefont
  {Chen}}]{chiang2017electrical}%
  \BibitemOpen
  \bibfield  {author} {\bibinfo {author} {\bibfnamefont {K.-H.}\ \bibnamefont
  {Chiang}}, \bibinfo {author} {\bibfnamefont {C.-L.}\ \bibnamefont {Lee}},
  \bibinfo {author} {\bibfnamefont {P.-Y.}\ \bibnamefont {Lai}},\ and\ \bibinfo
  {author} {\bibfnamefont {Y.-F.}\ \bibnamefont {Chen}},\ }\bibfield  {title}
  {\bibinfo {title} {Electrical autonomous brownian gyrator},\ }\href
  {https://doi.org/10.1103/PhysRevE.96.032123} {\bibfield  {journal} {\bibinfo
  {journal} {Phys. Rev. E}\ }\textbf {\bibinfo {volume} {96}},\ \bibinfo
  {pages} {032123} (\bibinfo {year} {2017})}\BibitemShut {NoStop}%
\bibitem [{\citenamefont {Gonzalez}\ \emph {et~al.}(2019)\citenamefont
  {Gonzalez}, \citenamefont {Neu},\ and\ \citenamefont
  {Teitsworth}}]{gonzalez2019experimental}%
  \BibitemOpen
  \bibfield  {author} {\bibinfo {author} {\bibfnamefont {J.~P.}\ \bibnamefont
  {Gonzalez}}, \bibinfo {author} {\bibfnamefont {J.~C.}\ \bibnamefont {Neu}},\
  and\ \bibinfo {author} {\bibfnamefont {S.~W.}\ \bibnamefont {Teitsworth}},\
  }\bibfield  {title} {\bibinfo {title} {Experimental metrics for detection of
  detailed balance violation},\ }\href
  {https://doi.org/10.1103/PhysRevE.99.022143} {\bibfield  {journal} {\bibinfo
  {journal} {Phys. Rev. E}\ }\textbf {\bibinfo {volume} {99}},\ \bibinfo
  {pages} {022143} (\bibinfo {year} {2019})}\BibitemShut {NoStop}%
\bibitem [{\citenamefont {Dotsenko}\ \emph {et~al.}(2013)\citenamefont
  {Dotsenko}, \citenamefont {Macio\l{}ek}, \citenamefont {Vasilyev},\ and\
  \citenamefont {Oshanin}}]{dotsenko2013two}%
  \BibitemOpen
  \bibfield  {author} {\bibinfo {author} {\bibfnamefont {V.}~\bibnamefont
  {Dotsenko}}, \bibinfo {author} {\bibfnamefont {A.}~\bibnamefont
  {Macio\l{}ek}}, \bibinfo {author} {\bibfnamefont {O.}~\bibnamefont
  {Vasilyev}},\ and\ \bibinfo {author} {\bibfnamefont {G.}~\bibnamefont
  {Oshanin}},\ }\bibfield  {title} {\bibinfo {title} {Two-temperature langevin
  dynamics in a parabolic potential},\ }\href
  {https://doi.org/10.1103/PhysRevE.87.062130} {\bibfield  {journal} {\bibinfo
  {journal} {Phys. Rev. E}\ }\textbf {\bibinfo {volume} {87}},\ \bibinfo
  {pages} {062130} (\bibinfo {year} {2013})}\BibitemShut {NoStop}%
\bibitem [{\citenamefont {Cerasoli}\ \emph {et~al.}(2018)\citenamefont
  {Cerasoli}, \citenamefont {Dotsenko}, \citenamefont {Oshanin},\ and\
  \citenamefont {Rondoni}}]{cerasoli2018asymmetry}%
  \BibitemOpen
  \bibfield  {author} {\bibinfo {author} {\bibfnamefont {S.}~\bibnamefont
  {Cerasoli}}, \bibinfo {author} {\bibfnamefont {V.}~\bibnamefont {Dotsenko}},
  \bibinfo {author} {\bibfnamefont {G.}~\bibnamefont {Oshanin}},\ and\ \bibinfo
  {author} {\bibfnamefont {L.}~\bibnamefont {Rondoni}},\ }\bibfield  {title}
  {\bibinfo {title} {Asymmetry relations and effective temperatures for biased
  brownian gyrators},\ }\href {https://doi.org/10.1103/PhysRevE.98.042149}
  {\bibfield  {journal} {\bibinfo  {journal} {Phys. Rev. E}\ }\textbf {\bibinfo
  {volume} {98}},\ \bibinfo {pages} {042149} (\bibinfo {year}
  {2018})}\BibitemShut {NoStop}%
\bibitem [{\citenamefont {Mancois}\ \emph {et~al.}(2018)\citenamefont
  {Mancois}, \citenamefont {Marcos}, \citenamefont {Viot},\ and\ \citenamefont
  {Wilkowski}}]{mancois2018two}%
  \BibitemOpen
  \bibfield  {author} {\bibinfo {author} {\bibfnamefont {V.}~\bibnamefont
  {Mancois}}, \bibinfo {author} {\bibfnamefont {B.}~\bibnamefont {Marcos}},
  \bibinfo {author} {\bibfnamefont {P.}~\bibnamefont {Viot}},\ and\ \bibinfo
  {author} {\bibfnamefont {D.}~\bibnamefont {Wilkowski}},\ }\bibfield  {title}
  {\bibinfo {title} {Two-temperature brownian dynamics of a particle in a
  confining potential},\ }\href {https://doi.org/10.1103/PhysRevE.97.052121}
  {\bibfield  {journal} {\bibinfo  {journal} {Phys. Rev. E}\ }\textbf {\bibinfo
  {volume} {97}},\ \bibinfo {pages} {052121} (\bibinfo {year}
  {2018})}\BibitemShut {NoStop}%
\bibitem [{\citenamefont {Nascimento}\ and\ \citenamefont
  {Morgado}(2019)}]{nascimento2019memory}%
  \BibitemOpen
  \bibfield  {author} {\bibinfo {author} {\bibfnamefont {E.}~\bibnamefont
  {Nascimento}}\ and\ \bibinfo {author} {\bibfnamefont {W.~A.}\ \bibnamefont
  {Morgado}},\ }\bibfield  {title} {\bibinfo {title} {Memory effects on
  two-dimensional overdamped brownian dynamics},\ }\href
  {https://doi.org/10.1088/1751-8121/ab5e2b} {\bibfield  {journal} {\bibinfo
  {journal} {J. Phys. A.}\ }\textbf {\bibinfo {volume} {53}},\ \bibinfo {pages}
  {065001} (\bibinfo {year} {2019})}\BibitemShut {NoStop}%
\bibitem [{\citenamefont {Asheichyk}\ and\ \citenamefont
  {Kr\"uger}(2019)}]{asheichyk2019using}%
  \BibitemOpen
  \bibfield  {author} {\bibinfo {author} {\bibfnamefont {K.}~\bibnamefont
  {Asheichyk}}\ and\ \bibinfo {author} {\bibfnamefont {M.}~\bibnamefont
  {Kr\"uger}},\ }\bibfield  {title} {\bibinfo {title} {Using the
  fluctuation-dissipation theorem for nonconservative forces},\ }\href
  {https://doi.org/10.1103/PhysRevResearch.1.033151} {\bibfield  {journal}
  {\bibinfo  {journal} {Phys. Rev. Research}\ }\textbf {\bibinfo {volume}
  {1}},\ \bibinfo {pages} {033151} (\bibinfo {year} {2019})}\BibitemShut
  {NoStop}%
\bibitem [{\citenamefont {Risken}(1996)}]{risken1996fokker}%
  \BibitemOpen
  \bibfield  {author} {\bibinfo {author} {\bibfnamefont {H.}~\bibnamefont
  {Risken}},\ }\href@noop {} {\emph {\bibinfo {title} {The Fokker-Planck
  Equation}}}\ (\bibinfo  {publisher} {Springer, Berlin},\ \bibinfo {year}
  {1996})\BibitemShut {NoStop}%
\bibitem [{\citenamefont {Weiss}\ \emph {et~al.}(2019)\citenamefont {Weiss},
  \citenamefont {Fox-Kemper}, \citenamefont {Mandal}, \citenamefont {Nelson},\
  and\ \citenamefont {Zia}}]{weiss2019nonequilibrium}%
  \BibitemOpen
  \bibfield  {author} {\bibinfo {author} {\bibfnamefont {J.~B.}\ \bibnamefont
  {Weiss}}, \bibinfo {author} {\bibfnamefont {B.}~\bibnamefont {Fox-Kemper}},
  \bibinfo {author} {\bibfnamefont {D.}~\bibnamefont {Mandal}}, \bibinfo
  {author} {\bibfnamefont {A.~D.}\ \bibnamefont {Nelson}},\ and\ \bibinfo
  {author} {\bibfnamefont {R.~K.}\ \bibnamefont {Zia}},\ }\bibfield  {title}
  {\bibinfo {title} {Nonequilibrium oscillations, probability angular momentum,
  and the climate system},\ }\href {https://doi.org/10.1007/s10955-019-02394-1}
  {\bibfield  {journal} {\bibinfo  {journal} {J. Stat. Phys.}\ }\textbf
  {\bibinfo {volume} {179}},\ \bibinfo {pages} {1} (\bibinfo {year}
  {2019})}\BibitemShut {NoStop}%
\bibitem [{\citenamefont {Sekimoto}(2010)}]{sekimoto2010stochastic}%
  \BibitemOpen
  \bibfield  {author} {\bibinfo {author} {\bibfnamefont {K.}~\bibnamefont
  {Sekimoto}},\ }\href@noop {} {\emph {\bibinfo {title} {Stochastic
  energetics}}}\ (\bibinfo  {publisher} {Springer-Verlag},\ \bibinfo {year}
  {2010})\BibitemShut {NoStop}%
\bibitem [{\citenamefont {Chun}\ and\ \citenamefont
  {Noh}(2015)}]{chun2015hidden}%
  \BibitemOpen
  \bibfield  {author} {\bibinfo {author} {\bibfnamefont {H.-M.}\ \bibnamefont
  {Chun}}\ and\ \bibinfo {author} {\bibfnamefont {J.~D.}\ \bibnamefont {Noh}},\
  }\bibfield  {title} {\bibinfo {title} {Hidden entropy production by fast
  variables},\ }\href {https://doi.org/10.1103/PhysRevE.91.052128} {\bibfield
  {journal} {\bibinfo  {journal} {Phys. Rev. E}\ }\textbf {\bibinfo {volume}
  {91}},\ \bibinfo {pages} {052128} (\bibinfo {year} {2015})}\BibitemShut
  {NoStop}%
\bibitem [{\citenamefont {Park}\ \emph {et~al.}(2016)\citenamefont {Park},
  \citenamefont {Chun},\ and\ \citenamefont {Noh}}]{park2016efficiency}%
  \BibitemOpen
  \bibfield  {author} {\bibinfo {author} {\bibfnamefont {J.-M.}\ \bibnamefont
  {Park}}, \bibinfo {author} {\bibfnamefont {H.-M.}\ \bibnamefont {Chun}},\
  and\ \bibinfo {author} {\bibfnamefont {J.~D.}\ \bibnamefont {Noh}},\
  }\bibfield  {title} {\bibinfo {title} {Efficiency at maximum power and
  efficiency fluctuations in a linear brownian heat-engine model},\ }\href
  {https://doi.org/10.1103/PhysRevE.94.012127} {\bibfield  {journal} {\bibinfo
  {journal} {Phys. Rev. E}\ }\textbf {\bibinfo {volume} {94}},\ \bibinfo
  {pages} {012127} (\bibinfo {year} {2016})}\BibitemShut {NoStop}%
\bibitem [{\citenamefont {Volpe}\ and\ \citenamefont
  {Petrov}(2006)}]{volpe2006torque}%
  \BibitemOpen
  \bibfield  {author} {\bibinfo {author} {\bibfnamefont {G.}~\bibnamefont
  {Volpe}}\ and\ \bibinfo {author} {\bibfnamefont {D.}~\bibnamefont {Petrov}},\
  }\bibfield  {title} {\bibinfo {title} {Torque detection using brownian
  fluctuations},\ }\href {https://doi.org/10.1103/PhysRevLett.97.210603}
  {\bibfield  {journal} {\bibinfo  {journal} {Phys. Rev. Lett.}\ }\textbf
  {\bibinfo {volume} {97}},\ \bibinfo {pages} {210603} (\bibinfo {year}
  {2006})}\BibitemShut {NoStop}%
\bibitem [{\citenamefont {Di~Leonardo}\ \emph {et~al.}(2007)\citenamefont
  {Di~Leonardo}, \citenamefont {Ruocco}, \citenamefont {Leach}, \citenamefont
  {Padgett}, \citenamefont {Wright}, \citenamefont {Girkin}, \citenamefont
  {Burnham},\ and\ \citenamefont {McGloin}}]{leonardo2007parametric}%
  \BibitemOpen
  \bibfield  {author} {\bibinfo {author} {\bibfnamefont {R.}~\bibnamefont
  {Di~Leonardo}}, \bibinfo {author} {\bibfnamefont {G.}~\bibnamefont {Ruocco}},
  \bibinfo {author} {\bibfnamefont {J.}~\bibnamefont {Leach}}, \bibinfo
  {author} {\bibfnamefont {M.~J.}\ \bibnamefont {Padgett}}, \bibinfo {author}
  {\bibfnamefont {A.~J.}\ \bibnamefont {Wright}}, \bibinfo {author}
  {\bibfnamefont {J.~M.}\ \bibnamefont {Girkin}}, \bibinfo {author}
  {\bibfnamefont {D.~R.}\ \bibnamefont {Burnham}},\ and\ \bibinfo {author}
  {\bibfnamefont {D.}~\bibnamefont {McGloin}},\ }\bibfield  {title} {\bibinfo
  {title} {Parametric resonance of optically trapped aerosols},\ }\href
  {https://doi.org/10.1103/PhysRevLett.99.010601} {\bibfield  {journal}
  {\bibinfo  {journal} {Phys. Rev. Lett.}\ }\textbf {\bibinfo {volume} {99}},\
  \bibinfo {pages} {010601} (\bibinfo {year} {2007})}\BibitemShut {NoStop}%
\bibitem [{\citenamefont {Kwon}\ \emph {et~al.}(2005)\citenamefont {Kwon},
  \citenamefont {Ao},\ and\ \citenamefont {Thouless}}]{kwon2005structure}%
  \BibitemOpen
  \bibfield  {author} {\bibinfo {author} {\bibfnamefont {C.}~\bibnamefont
  {Kwon}}, \bibinfo {author} {\bibfnamefont {P.}~\bibnamefont {Ao}},\ and\
  \bibinfo {author} {\bibfnamefont {D.~J.}\ \bibnamefont {Thouless}},\
  }\bibfield  {title} {\bibinfo {title} {Structure of stochastic dynamics near
  fixed points},\ }\href {https://doi.org/10.1073/pnas.0506347102} {\bibfield
  {journal} {\bibinfo  {journal} {Proc. Natl. Acad. Sci. U.S.A.}\ }\textbf
  {\bibinfo {volume} {102}},\ \bibinfo {pages} {13029} (\bibinfo {year}
  {2005})}\BibitemShut {NoStop}%
\bibitem [{\citenamefont {Kwon}\ \emph {et~al.}(2011)\citenamefont {Kwon},
  \citenamefont {Noh},\ and\ \citenamefont {Park}}]{kwon2011nonequilibrium}%
  \BibitemOpen
  \bibfield  {author} {\bibinfo {author} {\bibfnamefont {C.}~\bibnamefont
  {Kwon}}, \bibinfo {author} {\bibfnamefont {J.~D.}\ \bibnamefont {Noh}},\ and\
  \bibinfo {author} {\bibfnamefont {H.}~\bibnamefont {Park}},\ }\bibfield
  {title} {\bibinfo {title} {Nonequilibrium fluctuations for linear diffusion
  dynamics},\ }\href {https://doi.org/10.1103/PhysRevE.83.061145} {\bibfield
  {journal} {\bibinfo  {journal} {Phys. Rev. E}\ }\textbf {\bibinfo {volume}
  {83}},\ \bibinfo {pages} {061145} (\bibinfo {year} {2011})}\BibitemShut
  {NoStop}%
\bibitem [{\citenamefont {Hogg}\ \emph {et~al.}(2019)\citenamefont {Hogg},
  \citenamefont {McKean},\ and\ \citenamefont {Craig}}]{hogg2005introduction}%
  \BibitemOpen
  \bibfield  {author} {\bibinfo {author} {\bibfnamefont {R.~V.}\ \bibnamefont
  {Hogg}}, \bibinfo {author} {\bibfnamefont {J.}~\bibnamefont {McKean}},\ and\
  \bibinfo {author} {\bibfnamefont {A.~T.}\ \bibnamefont {Craig}},\ }\href@noop
  {} {\emph {\bibinfo {title} {Introduction to mathematical statistics}}}\
  (\bibinfo  {publisher} {Pearson Education, Boston},\ \bibinfo {year}
  {2019})\BibitemShut {NoStop}%
\bibitem [{\citenamefont {Weiss}(2003)}]{weiss2003coordinate}%
  \BibitemOpen
  \bibfield  {author} {\bibinfo {author} {\bibfnamefont {J.~B.}\ \bibnamefont
  {Weiss}},\ }\bibfield  {title} {\bibinfo {title} {Coordinate invariance in
  stochastic dynamical systems},\ }\href
  {https://doi.org/10.3402/tellusa.v55i3.12093} {\bibfield  {journal} {\bibinfo
   {journal} {Tellus Dyn. Meteorol. Oceanogr.}\ }\textbf {\bibinfo {volume}
  {55}},\ \bibinfo {pages} {208} (\bibinfo {year} {2003})}\BibitemShut
  {NoStop}%
\bibitem [{\citenamefont {Touchette}(2018)}]{touchette2018introduction}%
  \BibitemOpen
  \bibfield  {author} {\bibinfo {author} {\bibfnamefont {H.}~\bibnamefont
  {Touchette}},\ }\bibfield  {title} {\bibinfo {title} {Introduction to
  dynamical large deviations of markov processes},\ }\href
  {https://doi.org/10.1016/j.physa.2017.10.046} {\bibfield  {journal} {\bibinfo
   {journal} {Physica A}\ }\textbf {\bibinfo {volume} {504}},\ \bibinfo {pages}
  {5} (\bibinfo {year} {2018})}\BibitemShut {NoStop}%
\end{thebibliography}%

\end{document}